\documentclass[12pt]{iopart}

\pdfoutput=1 
\usepackage{graphicx}
\usepackage{amsfonts}
\usepackage{url}
\usepackage{cite}
\usepackage{color}
\usepackage{epstopdf}
\usepackage[breaklinks=true,colorlinks=true,linkcolor=blue,urlcolor=blue,citecolor=blue]{hyperref}

\usepackage{bm}

\def\Re{\mathrm{Re}}
\def\Im{\mathrm{Im}}

\def\e{\bm{e}}
\def\n{\bm{n}}

\def\x{\bm{x}}

\def\G{\bm{G}}

\def\C{\mathbb{C}}

\def\I{\mathbb{I}}

\def\R{\mathbb{R}}

\def\B{\mathcal{B}}

\def\pa{\partial\Omega}

\begin{document}

\title{Spectral properties of the Bloch-Torrey operator in three dimensions}

\author{Denis~S.~Grebenkov}
 \ead{denis.grebenkov@polytechnique.edu}
\address{
Laboratoire de Physique de la Mati\`{e}re Condens\'{e}e, \\ 
CNRS -- Ecole Polytechnique, Institut Polytechnique de Paris, 91120 Palaiseau, France}

\date{\today}

\begin{abstract}
We consider the Bloch-Torrey operator, $-\Delta + igx$, that governs
the time evolution of the transverse magnetization in diffusion
magnetic resonance imaging (dMRI).  Using the matrix formalism, we
compute numerically the eigenvalues and eigenfunctions of this
non-Hermitian operator for two bounded three-dimensional domains: a
sphere and a capped cylinder.  We study the dependence of its
eigenvalues and eigenfunctions on the parameter $g$ and on the shape
of the domain (its eventual symmetries and anisotropy).  In
particular, we show how an eigenfunction drastically changes its shape
when the associated eigenvalue crosses a branch (or exceptional)
point in the spectrum.  Potential implications of this behavior for
dMRI are discussed.
\end{abstract}

\pacs{02.50.-r, 05.40.-a, 02.70.Rr, 05.10.Gg}



\noindent{\it Keywords\/}: Bloch-Torrey operator, non-Hermitian operator, 
branch point, diffusion-weighted NMR, localization, pulsed-gradient spin-echo,
microstructure

\submitto{\JPA}

\maketitle

\section{Introduction}

Diffusion magnetic resonance imaging (dMRI) is a non-invasive
technique with multiple applications in medicine, neurosciences and
material sciences \cite{Callaghan,Price,LeBihan12,Novikov18}.  In a
typical setting, a static magnetic field $B_0$ is applied along the
$z$ axis to create the local magnetization of the nuclei (e.g.,
protons).  A radio-frequency (rf) $90^\circ$ pulse allows one to turn
the local magnetization into the transverse $xy$ plane, in which it
starts to precess around the $z$ axis.  If the static field $B_0$ is
superimposed with a spatially inhomogeneous magnetic field, the Larmor
frequency of each precessing nucleus depends on its spatial location,
allowing one to encode random trajectories of these nuclei that are
hindered by the environment and thus contain potentially exploitable
information on its structural properties.  Many theoretical and
numerical approaches have been developed to study this fundamental
problem (see reviews \cite{Axelrod01,Grebenkov07,Kiselev17} and
references therein).  The most common microscopic description of this
phenomenon relies on the Bloch-Torrey equation \cite{Torrey56} that
governs time evolution of the transverse magnetization $m(\x,t)$ of
the nuclei in a confining domain $\Omega
\subset \R^d$:
\begin{equation}  \label{eq:BTeq}
\partial_t m(\x,t) = D_0 \Delta m(\x,t) - i \gamma (\G(t) \cdot \x) m(\x,t) \quad (\x\in\Omega),
\end{equation}
where $D_0$ is the constant (self-)diffusion coefficient of the nuclei
(e.g., water molecules), $\Delta$ is the Laplace operator, $\gamma$ is
the gyromagnetic ratio of the nuclei, and $\G(t)$ is the gradient
profile of the applied magnetic field, which is set and controlled by
the experimental setup.
The Bloch-Torrey equation is usually complemented by the uniform
initial condition, $m(\x,0) = m_0 = 1/|\Omega|$, reflecting the
homogeneous excitation of the nuclei at time $t=0$ by the rf pulse in
the volume $|\Omega|$ of the confining domain $\Omega$.  The confining
microstructure is incorporated via an appropriate boundary condition.
A typical situation of an impenetrable inert surface $\pa$ is
described by Neumann boundary condition, $\partial_n
m(\x,t)\bigr|_{\pa} = 0$, stating that the magnetization flux across
the surface is zero, where $\partial_n$ is the normal derivative
oriented outwards the confining domain $\Omega$.  Surface relaxation
due to magnetic impurities on the boundary or nuclear exchange across
permeable membranes can also be described by modifying the boundary
condition \cite{Grebenkov10,Nguyen14}.  In addition, $T_1$ and $T_2$
bulk relaxation mechanisms can be included into Eq. (\ref{eq:BTeq}).
Since the transverse magnetization in any point $\x$ is too small to
be measured, only its integral over the confining domain (or a voxel)
is accessible in experiments:
\begin{equation}
S = \int\limits_{\Omega} d\x \, m(\x,t).
\end{equation}
This macroscopic signal that can be accessed as a function of the
gradient profile $\G(t)$, aggregates the microstructural features in a
very sophisticated way through the boundary condition to the
Bloch-Torrey equation (\ref{eq:BTeq}).
The imaginary unit $i$ in front of the last term of
Eq. (\ref{eq:BTeq}), which represents precession of the nuclei in the
transverse plane, makes this classical diffusion-reaction problem
challenging.  In fact, the differential operator governing time
evolution is not Hermitian that results in numerous unexpected
features such as the failure of perturbative approaches at high
gradients, localization near specific points on the boundary, or
branch points in the spectrum
\cite{Axelrod01,Grebenkov07,Moutal20}.

In order to understand the intricate relation between the
microstructure and the signal, one can focus on piecewise constant
gradient profiles and study the magnetization evolution during one
constant gradient pulse, i.e., to set $\G(t) = \G$.  Denoting by $x$
the coordinate axis in the direction of the gradient, one has $(\G
\cdot \x) = G x$, where $G = |\G|$ is the gradient amplitude, and $x$
is the projection of $\x$ onto the direction of $\G$.  Introducing the
Bloch-Torrey operator as
\begin{equation}  \label{eq:BT_def}
\B_g = - \Delta + i g x   \quad (g = \gamma G/D_0),
\end{equation}
one can formally solve the Bloch-Torrey equation as $m(\x,t) =
\exp(-D_0 \B_g t) m_0$.  In other words, the effect of a constant
gradient pulse is represented by the evolution operator $\exp(-D_0
\B_g t)$.  One can also deal with more sophisticated gradient
profiles by representing them as a sequence of constant gradient
pulses and combining the corresponding evolution operators
\cite{Barzykin98,Barzykin99,Grebenkov08b,Ozarslan09}.  For instance,
in a standard Stejskal-Tanner pulsed-gradient spin echo (PGSE)
sequence with two rectangular gradient pulses of duration $\delta$ and
opposite directions \cite{Tanner68}, the signal can be written as
\begin{equation}   \label{eq:S_matrix}
S = \int\limits_{\Omega} d\x \, \biggl(e^{-D_0 \delta \B_{-g}} e^{-D_0 \delta \B_{g}} \frac{1}{|\Omega|} \biggr) ,
\end{equation}
where $e^{-D_0 \delta\B_g}$ represents the evolution from the initial
uniform magnetization $m_0 = 1/|\Omega|$ during the first gradient
pulse, and $e^{-D_0 \delta \B_{-g}}$ describes the evolution during
the second gradient pulse with the opposite direction (for simplicity,
we assumed here that the second pulse starts immediately after the
first one).
When the Bloch-Torrey operator has a discrete spectrum, one can use
its eigenvalues $\lambda_j^{(g)}$ and eigenfunctions $v_j^{(g)}$
(enumerated by $j = 1,2,\ldots$) to represent the above signal as
\cite{deSwiet94,Grebenkov14,Herberthson17,Moutal19}
\begin{equation}  \label{eq:S_exact}
S = \sum\limits_{j,j'=1}^\infty C_{j,j'}^{(g)} \, e^{-D_0 \delta (\lambda_j^{(-g)} + \lambda_{j'}^{(g)})}   ,
\end{equation}
where the coefficients
\begin{equation}  \label{eq:C_def}
\fl \qquad
C_{j,j'}^{(g)} = \frac{1}{|\Omega|} \left(\int\limits_\Omega d\x \, v_j^{(-g)}(\x)\right)
\left(\int\limits_\Omega d\x \, v_j^{(-g)}(\x) \, v_{j'}^{(g)}(\x)\right)
\left(\int\limits_\Omega d\x \, v_{j'}^{(g)}(\x)\right)
\end{equation}
characterize the overlap between two eigenfunctions $v_j^{(-g)}$ and
$v_{j'}^{(g)}$, and their projections onto a constant.  As a
consequence, the macroscopic signal $S$ and its dependence on the
microstructure are fully determined by the spectral properties of the
Bloch-Torrey operator.  Moreover, when gradient pulses are long and/or
strong enough such that $D_0 \delta \Re\{\lambda_1^{(g)}\} \gg 1$, the
above expansion can be truncated to few terms, yielding a practical
approximation for the signal, as discussed below.

The seminal paper by Stoller, Happer and Dyson provided the first
thorough analysis of the Bloch-Torrey operator in one dimension (for
an interval and a half-line) \cite{Stoller91}.  In particular, they
showed that the spectrum is discrete, while the eigenvalues
$\lambda_{k}^{(g)}$ of the Bloch-Torrey operator $\B_g$ behave as
$\lambda_k^{(g)} \propto g^{2/3} \propto G^{2/3}$ at large $G$, that
results in the specific long-time decay of the signal, $\ln S \propto
G^{2/3} t$, with unexpected ``anomalous'' dependence $G^{2/3}$ on the
gradient.  This behavior is drastically different from the common
quadratic dependence, $\ln S \propto G^2$, that appears at small
gradients in both slow-diffusion and motional-narrowing regimes
\cite{Robertson66,Neuman74,Grebenkov07}.  The spectral analysis was
later extended to different classes of confining domains, including an
arbitrary array of permeable intervals
\cite{Grebenkov14,Grebenkov17,Moutal19b}, a disk and a sphere
\cite{deSwiet94}, bounded planar domains
\cite{Herberthson17,Grebenkov18}, the exterior of compact domains
\cite{Almog18,Almog19}, and periodic domains
\cite{Moutal20b,Grebenkov21}.  Most focus was on the large-$G$
asymptotic behavior of the eigenvalues and on the localization of
eigenfunctions.  Moreover, the whole structure of the spectrum,
including the existence of branch points (also known as exceptional
or diabolic points), was investigated \cite{Moutal22}.  The existence
of branch points is a peculiar feature of non-Hermitian operators
(see,
e.g. \cite{Berry04,Heiss04,Seyranian05,Kirillov05,Rubinstein07,Cartarius07,Cejnar07,Klaiman08,Chang09,Ceci11,Shapiro17,Grosfjeld19}
and references therein).  The ``anomalous'' $G^{2/3}$-dependence of
$\ln S$ was first confirmed experimentally by H\"urlimann {\it et al.}
for diffusion of water molecules between two parallel planes
\cite{Hurlimann95}, and later for gas diffusion in cylindrical
phantoms \cite{Moutal19}.  Experimental evidence for the localization
regime in biological samples was reported \cite{Williamson19}.

In this paper, we extend the recent analysis from Ref. \cite{Moutal22}
that was focused on planar domains, into three dimensions.  First, we
uncover the behavior of eigenvalues and eigenfunctions of the
Bloch-Torrey operator for a sphere.  While the spherical confinement
is one of the most archetypical models in this field, a systematic
study of the spectral properties of $\B_g$ in this setting is still
missing.  In particular, we analyze the dependence of eigenvalues on
the gradient and reveal the existence of branch points in the spectrum
of $\B_g$ for this domain.  We also discuss one-mode and two-modes
approximations of the macroscopic signal.  Second, we analyze the
spectrum of the Bloch-Torrey operator for a capped cylinder that
exhibits structural anisotropy.  We show how the structure of the
spectrum depends on the gradient direction, in particular, how the
branch points can be tuned experimentally.  The structure of the
underlying eigenfunctions is discussed.

The paper is organized as follows.  In Sec. \ref{sec:reminder}, we
recall some basic spectral properties of the Bloch-Torrey operator
$\B_g$.  Section \ref{sec:sphere} presents the detailed analysis for
the case of a sphere; in particular, we discuss the dependence of the
eigenvalues on $g$, the branch points in the spectrum, and the drastic
change of eigenfunctions at these points.  In turn,
Sec. \ref{sec:capped} focuses on a capped cylinder that exhibits
anisotropy and allows us to reveal its impact onto the spectrum.
Section \ref{sec:conclusion} concludes the paper by summarizing the
main results and presenting their practical implications in diffusion
MRI.  Appendices contain technical discussions such as the description
of the numerical procedure for constructing the spectrum of the
Bloch-Torrey operator by using the matrix formalism
(\ref{sec:numerics}), the matrix elements for a sphere
(\ref{sec:Asphere}) and for a capped cylinder (\ref{sec:Acapped}), as
well as a simple orthogonalization procedure for eigenfunctions with
degenerate eigenvalues (\ref{sec:A_ortho}).

\section{Summary of basic spectral properties}
\label{sec:reminder}

For a given bounded domain $\Omega$ with a smooth boundary $\pa$, we
are interested in the spectral properties of the Bloch-Torrey operator
$\B_g$ defined in Eq. (\ref{eq:BT_def}).  As the parameter $g = \gamma
G/D_0$ is determined by the amplitude of the gradient used in
diffusion MRI, we mainly focus on positive values $g \geq 0$.

In this section, we remind basic spectral properties of the
non-Hermitian Bloch-Torrey operator for $g > 0$ (see further
discussion in \cite{Moutal22} and references therein).
As $igx$ is a bounded perturbation of the (unbounded) Laplace
operator, the spectrum is discrete, i.e., there is an infinite
sequence of eigenvalues $\lambda_j^{(g)}$ and eigenfunctions
$v_j^{(g)}(\x)$ satisfying
\begin{equation}  \label{eq:eigen_def}
\B_g \, v_j^{(g)}(\x) = \lambda_j^{(g)} \, v_j^{(g)}(\x)  \quad (\x\in \Omega),  
\qquad \partial_n v_j^{(g)}(\x) = 0 \quad (\x\in\pa).
\end{equation}
The eigenfunctions are in general complex-valued.

Since the Bloch-Torrey operator is not Hermitian for $g > 0$, the
standard scalar product in $L_2(\Omega)$, $(u,v) =
\int\nolimits_\Omega d\x \, u^*(\x) v(\x)$, is replaced by a bilinear
form $\langle u,v\rangle = \int\nolimits_\Omega d\x \, u(\x) v(\x)$.
In particular, the eigenfunctions $\{v_j^{(g)}\}$ are in general not
orthogonal to each other, $(v_j^{(g)}, v_{j'}^{(g)}) \ne 0$, as it
would be for Hermitian operators (e.g., for $\B_0$).  In turn, one can
easily show by the Green's formula that
\begin{equation}  \label{eq:vj_orthogonality}
\bigl(\lambda_j^{(g)} - \lambda_{j'}^{(g)}\bigr) \langle v_j^{(g)}, v_{j'}^{(g)}\rangle = 0  ,
\end{equation}
so that if the eigenvalues $\lambda_j^{(g)}$ and $\lambda_{j'}^{(g)}$
are not equal, then $\langle v_j^{(g)}, v_{j'}^{(g)}\rangle = 0$.  It
is worth stressing that $\langle v,v\rangle$ is not a norm of $v$; in
particular, there exist special values of $g$ (so-called branch
points, see below), at which $\langle v_j^{(g)}, v_{j}^{(g)}\rangle =
\int\nolimits_\Omega d\x \, [v_j^{(g)}]^2 = 0$.  In general, however,
this integral is not zero, and we normalize the eigenfunctions to have
\begin{equation}   \label{eq:vj_norm}
\langle v_j^{(g)}, v_{j}^{(g)}\rangle = 1. 
\end{equation}
This condition fixes the normalization up to a factor $\pm 1$.

The eigenvalues are in general complex-valued, with positive real
parts that accumulate at $+\infty$.  It is therefore convenient to
order the eigenvalues according to their increasing real parts.
However, we will adopt a different ordering procedure.  In fact, the
eigenvalues $\lambda_j^{(g)}$ can be understood as different branches
in the complex plane $\C$ of a multi-valued function $\lambda(g)$
defined implicitly as the solution of the transcendental equation
$\det(\B_g - \lambda(g) \mathcal{I}) = 0$ for any fixed $g$, where
$\mathcal{I}$ is the identity operator (see \cite{Moutal22} for more
details).  This formal definition resembles the practical procedure
for computing the eigenvalues when the Bloch-Torrey operator $\B_g$ is
represented by an infinite-dimensional matrix, which is then truncated
and diagonalized numerically (see \ref{sec:numerics}).  The eigenvalue
branches $\lambda_j^{(g)}$ can merge and split at branch points
but, apart from these points, they are smooth functions of $g$.  We
use this property to order the eigenvalues $\lambda_j^{(g)}$ according
to the increasing order of Laplacian eigenvalues $\lambda_j^{(0)}$.
In other words, one first orders the eigenvalues at $g = 0$ and then
preserves their order by continuity of branches as $g$ increases.  At
each branch point, the order of merged eigenvalues is lost but they
can be re-ordered in any convenient way.  This ordering procedure does
not ensure an increasing order of $\Re\{\lambda_j^{(g)}\}$ for any $g$
but it facilitates the visualization and interpretation of the
spectrum.  Most importantly, the associated eigenfunctions
$v_j^{(g)}(\x)$ also change smoothly with $g$ and preserve their
symmetries, except for branch points (see below).

\section{Bloch-Torrey operator for a sphere}
\label{sec:sphere}

We consider restricted diffusion inside a sphere of radius $R$ with
reflecting boundary and apply the gradient along the $z$ axis: $\G = G
\bm{e}_z$.  The Bloch-Torrey operator can be written in spherical
coordinates $(r,\theta,\phi)$ as
\begin{equation}   \label{eq:BT_sphere}
\B_g^z = - \biggl(\partial_r^2 + \frac{2}{r}\partial_r + \frac{1}{r^2} \partial_{\xi}(1-\xi^2)\partial_\xi 
+ \frac{1}{r^2(1-\xi^2)} \partial_{\phi}^2\biggr) + ig r \xi ,
\end{equation}
where $\xi = \cos\theta$.  Since the gradient operator does not depend
on the azimuthal angle $\phi$ and the initial transverse magnetization
is uniform, the considered problem is axisymmetric with respect to the
$z$ axis.  In other words, the Bloch-Torrey equation does not change
the uniformity with respect to $\phi$, i.e., the transverse
magnetization remains independent of $\phi$.  For this reason, one
often considers the {\it reduced} Bloch-Torrey operator without the
azimuthal part:
\begin{equation}  \label{eq:BTreduced_sphere}
\hat{\B}_g = -\biggl(\partial_r^2 + \frac{2}{r}\partial_r + \frac{1}{r^2} \partial_{\xi}(1-\xi^2)\partial_\xi\biggr) + ig r \xi .
\end{equation}
In fact, most former studies were focused in this operator and its
matrix representation on the basis of the Laplace operator
\cite{Callaghan97,Barzykin98,Barzykin99,Grebenkov07,Grebenkov08,Grebenkov08b,Grebenkov10}.  
In turn, the full operator $\B_g^z$ in Eq. (\ref{eq:BT_sphere}) is
needed to deal with gradient pulses in different directions or with
inhomogeneous initial magnetization.  The related extension of the
matrix formalism was introduced in \cite{Ozarslan09}.  In
\ref{sec:Asphere}, we recall the matrix elements for constructing the
eigenvalues and eigenfunctions of both operators.
Similarly, one can introduce the Bloch-Torrey operators $\B_g^x$ and
$\B_g^y$ when the gradient is applied along $x$ and $y$ coordinates,
respectively.  Even though these two operators have different matrix
representations (see \ref{sec:Asphere}), the rotational invariance of
the sphere ensures that the spectra of the three operators $\B_g^x$,
$\B_g^y$ and $\B_g^z$ are identical.  In turn, their eigenfunctions
can be matched by an appropriate rotation of spherical coordinates
(i.e., by choosing the spherical coordinates with the $z$ axis aligned
with the desired gradient direction).  For this reason, we focus on
the operator $\B_g^z$ in the following and compare its spectral
properties to those of the reduced operator $\hat{\B}_g$.

When there is no gradient ($g = 0$), the eigenbasis of the (negative)
Laplace operator $\B_0^z = -\Delta$ is fairly well known; in particular,
the separation of variables yields the Laplacian eigenfunctions
$u_{nkm}(r,\theta,\phi) \propto j_n(\alpha_{nk} r/R) P_n^m(\cos\theta)
e^{im\phi}$, where $j_n(z)$ is the spherical Bessel function of the
first kind, $P_n^m(z)$ is the associated Legendre polynomial, and
$\alpha_{nk}$ are the positive zeros of the derivative $j'_n(z)$
ensuring the Neumann boundary condition.  Here each Laplacian
eigenfunction is parameterized by a triple index $nkm$ that reflects
its symmetries, with $n = 0,1,2,\ldots$ being the order of $j_n(z)$,
$k = 0,1,2,\ldots$ enumerating the zeros $\alpha_{nk}$, and $m =
-n,-n+1,\ldots,n$.  The associated eigenvalues $\lambda_{nkm} =
\alpha_{nk}^2/R^2$ do not depend on $m$ and are thus $(2n+1)$ times
degenerate.  Writing these eigenvalues in an increasing order (see
Table \ref{tab:lam}), we use the position $j$ of each eigenvalue in
the sequence to enumerate the branches $\lambda_j^{(g)}$.
Some ambiguities in the eigenvalue ordering procedure caused by the
degeneracy of the Laplacian eigenvalues can be fixed manually.

\begin{table}
\begin{center}
\begin{eqnarray*}
&&
\begin{array}{c| c| c c c| c c c c c}
j   & {\bf 1}  & {\bf 2}  &   3    &  4  &    5   &  6  & {\bf 7}  &   8    &  9 \\  \hline
nkm & 000 & 100 & 10(-1) & 101 & 20(-1) & 201 & 200 & 20(-2) & 202 \\   
\lambda_j^{(0)} & 0 & 4.33 & 4.33 & 4.33 & 11.17 & 11.17 & 11.17 & 11.17 & 11.17   \\ \end{array} \\
&&
\begin{array}{c| c| c c c c c c c}
j   & {\bf 10}  & {\bf 11} &   12   & 13  &  14    & 15  &  16    &  17 \\  \hline
nkm & 010       & 300 & 30(-2) & 302 & 30(-1) & 301 & 30(-3) & 303 \\   
\lambda_j^{(0)} & 20.19 & 20.38 & 20.38 & 20.38 & 20.38 & 20.38 & 20.38 & 20.38  \\ \end{array}
\end{eqnarray*}
\end{center}
\caption{
First 17 eigenvalues $\lambda_{nkm} = \lambda_j^{(0)}$ of the
(negative) Laplace operator, $\B_0 = -\Delta$, in the unit sphere with
reflecting boundary.  The position $j$ of the eigenvalue
$\lambda_j^{(0)}$ in the ordered sequence is used to enumerate the
branch $\lambda_j^{(g)}$ for $g \ne 0$.  Bold font highlights the
indices of eigenfunctions that are axisymmetric (with $m = 0$).}
\label{tab:lam}
\end{table}

\begin{figure}
\begin{center}
\includegraphics[width=120mm]{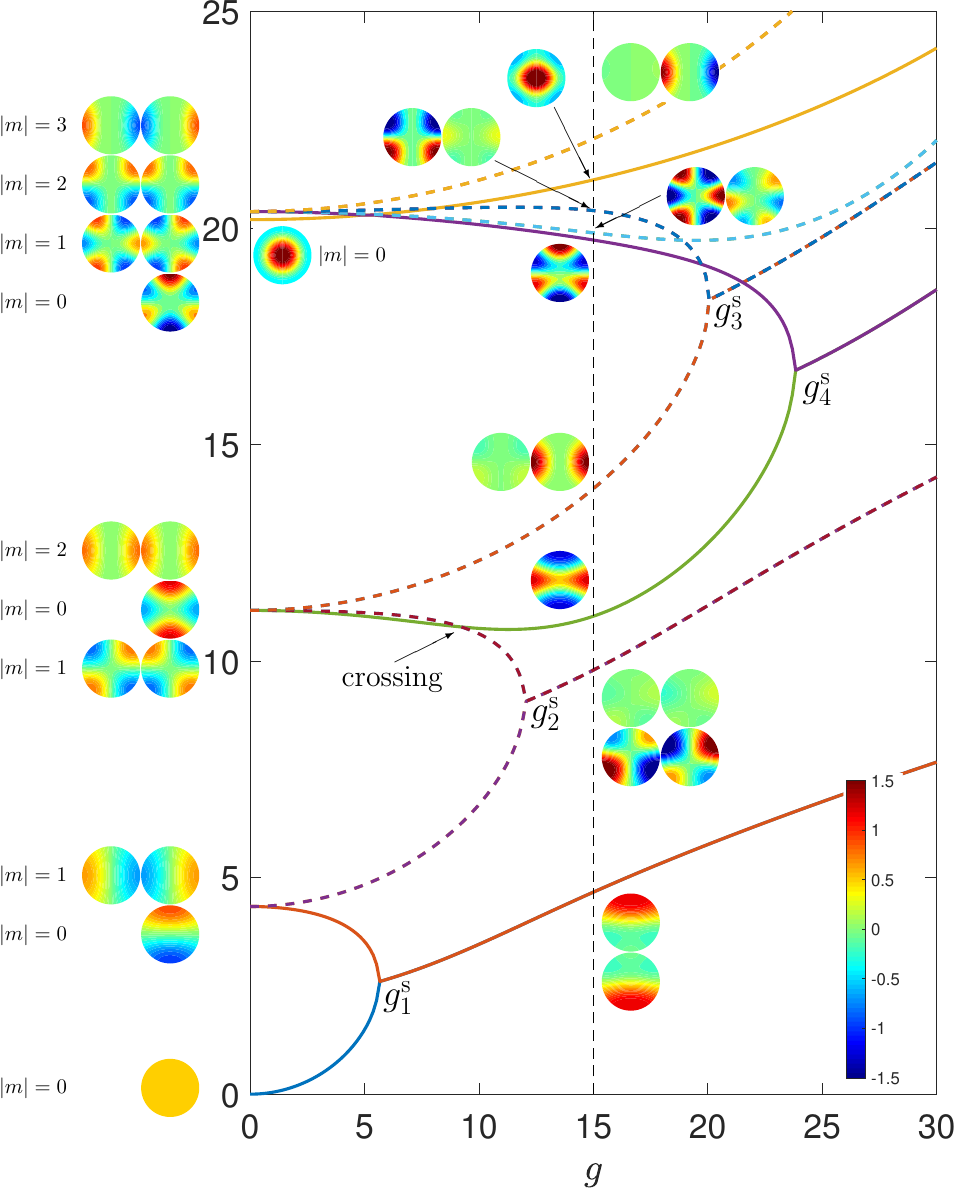} 
\end{center}
\caption{
Real part of the first 17 eigenvalues $\lambda_j^{(g)}$ of the
Bloch-Torrey operator $\B_g^z$ for the unit sphere ($R = 1$).  Dashed
lines indicate the eigenvalues that do not contribute to the
macroscopic signal and thus do not appear in the spectrum of the
reduced Bloch-Torrey operator $\hat{\B}_g$.
Colored snapshots show the $xz$ projection of the real part of the
corresponding eigenfunction, evaluated at $g = 0$ (on the left) and at
$g = 15$ (near vertical dashed line).  Color indicates changes of
$\Re\{v_j^{(g)}\}$ from $-1.5$ (dark blue) to $1.5$ (dark red), with
the colorbar shown at right bottom, being the same for all snapshots.
The values of $|m|$ determining the dependence $e^{im\phi}$ on the
azimuthal angle $\phi$ are shown on the left.  Four branch points
are seen: $g_1^{\rm s}\approx 5.622$, $g_2^{\rm s} \approx 12.1$,
$g_3^{\rm s} \approx 20.1$, and $g_4^{\rm s} \approx 23.84$.  The
eigenvalues and eigenfunctions were constructed via the matrix
formalism (see \ref{sec:numerics}), in which the matrices were
truncated at 333.  }
\label{fig:sphere_spectrum}
\end{figure}

\subsection{Eigenvalues}

Figure \ref{fig:sphere_spectrum} summarizes the spectral properties of
the Bloch-Torrey operator $\B_g^z$ for the unit sphere ($R = 1$).  One
sees the real part of first 17 eigenvalues $\lambda_j^{(g)}$ as
functions of $g$.  At $g = 0$, one retrieves the Laplacian eigenvalues
$\lambda_{nkm}$ with their degeneracies.  For instance, three branches
$\lambda_2^{(g)}$, $\lambda_3^{(g)}$ and $\lambda_4^{(g)}$ start from
$4.33$ at $g = 0$ but two of them coincide for all $g$,
$\lambda_3^{(g)} \equiv \lambda_4^{(g)}$, resulting in a single upper
curve.  Similarly, five branches $\lambda_5^{(g)}, \ldots,
\lambda_9^{(g)}$ start from $11.17$ at $g = 0$, but $\lambda_5^{(g)}
\equiv \lambda_6^{(g)}$ for all $g$ result in a single lower curve,
and $\lambda_8^{(g)} \equiv \lambda_9^{(g)}$ result in a single upper
curve.  These preserved degeneracies are related to the fact that the
Bloch-Torrey operator $\B_g^z$ does not affect the azimuthal angle.
In fact, as $P_n^{-m}(x) = (-1)^m P_n^m(x)$, two Laplacian
eigenfunctions corresponding to $+m$ and $-m$ exhibit the identical
dependence on $r$ and $\theta$ and therefore remain indistinguishable
even in the presence of the applied gradient along $z$ coordinate.  As
a consequence, the dependence $e^{im\phi}$ of the Laplacian
eigenfunctions on the angle $\phi$ is preserved for the eigenfunctions
of the Bloch-Torrey operator $\B_g^z$.  As the integral of the
eigenfunctions containing the factor $e^{im\phi}$ with $m \ne 0$ over
the sphere $\Omega$ vanishes, they do not contribute to the
macroscopic signal.  The related eigenvalues are shown by dashed
lines.  In turn, the eigenvalues shown by solid lines correspond to
the eigenfunctions $v_j^{(g)}$ that inherited their independence of
$\phi$ from the Laplacian eigenfunctions $u_{nk0}$ and thus do
contribute to the signal.  Expectedly, these eigenvalues could be
directly obtained by diagonalizing the reduced Bloch-Torrey operator
$\hat{\B}_g$.  In other words, the difference between the spectra of
the operators $\B_g^z$ and $\hat{\B}_g$ is the presence of additional
eigenvalues (shown by dashed lines) in the former case.

The rotation invariance of the sphere implies the PT symmetry of the
Bloch-Torrey operator \cite{Moiseyev11,El-Ganainy18}.  As a
consequence, its eigenvalues are either real, or form complex-conjugate
pairs (see \cite{Moutal19,Moutal22} for further discussions).  This
general property is confirmed on Fig. \ref{fig:sphere_spectrum}.
Moreover, one can observe four branch (or exceptional) points
$g_i^{\rm s}$, at which real eigenvalues merge to become
complex-conjugate pairs: $g_1^{\rm s}\approx 5.622$, $g_2^{\rm s}
\approx 12.1$, $g_3^{\rm s} \approx 20.1$, and $g_4^{\rm s} \approx 23.84$.
Note that the branch points $g_1^{\rm s}$ and $g_4^{\rm s}$ are of
order 2 (i.e., two simple eigenvalues merge here), while the branch
points $g_2^{\rm s}$ and $g_3^{\rm s}$ are of order 4 (two pairs of
twice degenerate eigenvalues merge).  To our knowledge, this is the
first observation of a branch point of order 4 for the Bloch-Torrey
operator (the previous studies \cite{Stoller91,Moutal22} revealed only
branch points of order 2).  As discussed earlier, twice degenerate
eigenvalues correspond to the eigenfunctions that do not contribute to
the signal.  In particular, the reduced Bloch-Torrey operator
$\hat{\B}_g$ seems to possess only branch points of order $2$.  We
also note that the preserved dependence of eigenfunctions on $\phi$
via $e^{im\phi}$ implies a simple branching rule: only the branches
containing at $g = 0$ the Laplacian eigenvalues $\lambda_{nkm}$ with
the same $|m|$ can merge.  For instance, the branches
$\lambda_1^{(g)}$ and $\lambda_2^{(g)}$ corresponding to
$\lambda_{000}$ and $\lambda_{100}$ (with $m = 0$) merge at $g_1^{\rm
s}$; the branches $\lambda_3^{(g)}, \ldots \lambda_6^{(g)}$
corresponding to $\lambda_{10(-1)}$, $\lambda_{101}$,
$\lambda_{20(-1)}$, $\lambda_{201}$ (with $|m|=1$) merge at $g_2^{\rm
s}$, and so on.  Note that our numerical study did not reveal
branch points of other orders except $2$ and $4$.  We expect that
their existence is unlikely but a mathematical proof of this statement
remains an open problem.
We also stress that branch points should be distinguished from
``crossing'' points, at which two (or more) eigenvalues cross, without
changing their properties.  For instance, the pair of real eigenvalues
$\lambda_5^{(g)} \equiv \lambda_6^{(g)}$ crosses a single real
eigenvalue $\lambda_7^{(g)}$ at $g \approx 9.3$.  Three corresponding
eigenfunctions form an orthogonal basis of the subspace of dimension
3.  In contrast, one (or more) eigenfunction disappears at the
branch point (see further discussion in \cite{Moutal22}).

\subsection{Eigenfunctions}

Figure \ref{fig:sphere_spectrum} also presents the $xz$ projections of
the real part of the first 17 eigenfunctions of the Bloch-Torrey
operator $\B_g^z$ at $g = 0$ and $g = 15$.  These snapshots help to
visualize how the geometric structure of each eigenfunction changes
with $g$.  As discussed earlier, the Laplacian eigenfunctions
$u_{nkm}$ and $u_{nk(-m)}$ exhibit the same dependence on $\phi$ and
thus keep this property in the presence of the gradient along the $z$
axis, as confirmed by snapshots at $g = 15$.  Note that the $xz$
projection of some eigenfunctions is close to $0$ (green color); in
fact, such an eigenfunction should be orthogonal to its pair and thus
exhibit most variations in other projections.  One also sees how the
symmetries of the first six eigenfunctions change after the branch
point.

\begin{figure}
\begin{center}
\includegraphics[width=\textwidth]{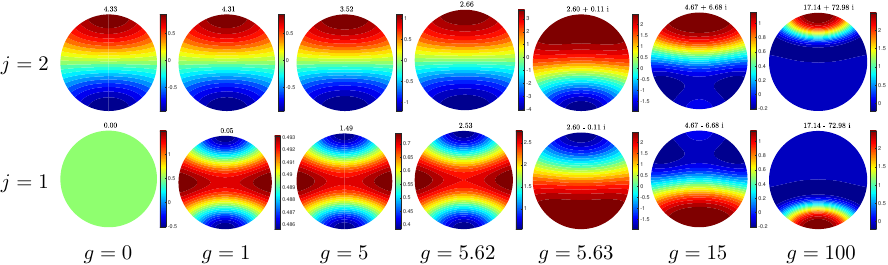} 
\end{center}
\caption{
$xz$ projection of the real part of the eigenfunctions $v_1^{(g)}$
(bottom row) and $v_2^{(g)}$ (top row) of the Bloch-Torrey operator
$\B_g^z$ for the unit sphere for several values of $g$.  The
associated eigenvalue is indicated on the top of each plot.  The
branch point is $g_1^{\rm s} \approx 5.622$.  Note that color range
changes between different panels.}
\label{fig:sphere_v1_branch}
\end{figure}

Let us inspect this change in more detail.  Figure
\ref{fig:sphere_v1_branch} illustrates the drastic change in the shape
of the eigenfunctions $v_1^{(g)}$ and $v_2^{(g)}$ when $g$ crosses the
branch point $g_1^{\rm s} \approx 5.622$.  We first consider the
eigenfunction $v_1^{(g)}$ (bottom row).  The uniform property of
$v_1^{(0)}$ is immediately broken for any $g > 0$, as confirmed by the
second panel showing $v_1^{(g)}$ at $g = 1$.  A similar geometric
pattern of $v_1^{(g)}$ was observed for even small $g$ (not shown).
It is worth noting, however, that $v_1^{(1)}$ varies from $0.489$ to
$0.493$ and thus remains very close to a constant (as $v_1^{(0)}$).
As $g$ increases up to $g_1^{\rm s}$, the shape of $v_1^{(g)}$ remains
visually unchanged but its variations grow rapidly.  This is the
consequence of the normalization by $\langle v_1^{(g)},
v_1^{(g)}\rangle^{-1/2}$.  In fact, as $g$ approaches the branch
point $g_1^{\rm s}$, $\langle v_1^{(g)},v_1^{(g)}\rangle$ vanishes and
thus the normalization factor diverges, as discussed in
\cite{Moutal22}.  At $g = 5.63 > g_1^{\rm s}$, the shape of the
eigenfunction $v_1^{(g)}$ has drastically changed and started to
exhibit variations along the $z$ axis, as imposed by the applied
gradient.  Further increase of $g$ does not change this symmetry but
enhances the localization of the eigenfunction on the South pole.

A similar behavior is observed for the second eigenfunction
$v_2^{(g)}$: its shape, inherited from the Laplacian eigenfunction
$u_{100}$, is preserved for $g < g_1^{\rm s}$ and then drastically
changes to another shape exhibiting variations along $z$ axis.
Moreover, as the eigenvalues $\lambda_1^{(g)}$ and $\lambda_2^{(g)}$
form a complex conjugate pair for $g > g_1^{\rm s}$, the associated
eigenfunctions exhibit the reflection symmetry: $v_2^{(g)}(\x) =
[v_1^{(g)}(R_z \x)]^*$, where $R_z$ is the reflection with respect to
the $xy$ plane (i.e., $z$ is replaced by $-z$).  Finally, a similar
behavior is observed (but not shown here) for other eigenfunctions
that drastically change their shapes at the branch point of their
eigenvalues (e.g., compare the eigenfunctions
$v_3^{(g)},\ldots,v_6^{(g)}$ shown in Fig. \ref{fig:sphere_spectrum}
at $g = 0$ and $g = 15$).

\subsection{Macroscopic signal}

When the duration $\delta$ of the gradient pulses is sufficiently
long, only few eigenmodes with small $\Re\{\lambda_j^{(g)}\}$ do
contribute to the signal.  The structure of the spectrum shown in
Fig. \ref{fig:sphere_spectrum} suggests to keep only the first two
eigenmodes in the spectral expansion (\ref{eq:S_exact}):
\begin{equation}  \label{eq:Sapprox}
\fl  \qquad
S \approx e^{-2D_0 \delta \lambda_1^{(g)}} \biggl[C_{1,1}^{(g)} + 2\Re\{C_{1,2}^{(g)}\} e^{-D_0\delta(\lambda_2^{(g)}-\lambda_1^{(g)})}
+ C_{2,2}^{(g)} e^{-2D_0\delta(\lambda_2^{(g)}-\lambda_1^{(g)})}\biggr] ,
\end{equation}
with the coefficients $C_{j,j'}^{(g)}$ given by Eq. (\ref{eq:C_def});
note that we used the property $\B_{-g} = \B_g^*$ that implies
$C_{j',j}^{(g)} = [C_{j,j'}^{(g)}]^*$.
One can distinguish two scenarios according to whether the eigenvalues
$\lambda_1^{(g)}$ and $\lambda_2^{(g)}$ are real or complex.

(i) When $0< g < g_1^{\rm s}$, the eigenvalues $\lambda_1^{(g)}$ and
$\lambda_2^{(g)}$ are real and simple.  If
$D_0\delta(\lambda_2^{(g)}-\lambda_1^{(g)}) \gg 1$, the last two terms
in Eq. (\ref{eq:Sapprox}) can be neglected, yielding the one-mode
approximation for the signal,
\begin{equation}  \label{eq:Sone}
S \approx S_{\rm one} = C_{1,1}^{(g)} \, e^{-2D_0 \delta \lambda_1^{(g)}} .
\end{equation}
Note that this approximation is not valid when $g$ is close to the
branch point $g_1^{\rm s}$.

(ii) When $g > g_1^{\rm s}$, the eigenvalue $\lambda_1^{(g)}$ is
complex and paired with $\lambda_2^{(g)} = [\lambda_1^{(g)}]^*$.  As a
consequence, one has $C_{2,2}^{(g)} = C_{1,1}^{(g)}$ so that
Eq. (\ref{eq:Sapprox}) can be written as
\begin{equation}   \label{eq:Stwo}
S \approx S_{\rm two}  = 2 e^{-2D_0 \delta\, \Re\{\lambda_1^{(g)}\} } \biggl[C_{1,1}^{(g)}
+ \Re\bigl\{ C_{1,2}^{(g)} \, e^{2iD_0 \delta\, \Im\{\lambda_1^{(g)}\}} \bigr\}\biggr].
\end{equation}

\begin{figure}
\begin{center}
\includegraphics[width=100mm]{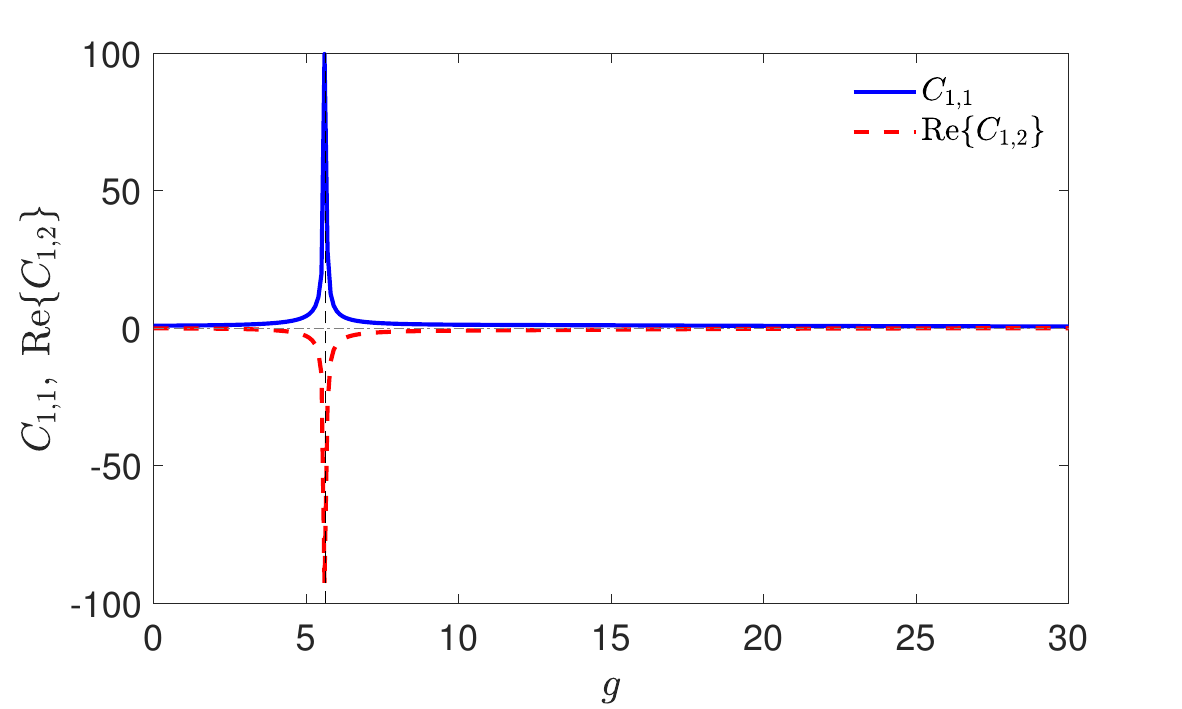} 
\end{center}
\caption{
The coefficients $C_{1,1}^{(g)}$ and $\Re\{C_{1,2}^{(g)}\}$ from
Eq. (\ref{eq:C_def}) characterizing the contributions of the first two
eigenfunctions to the signal for the unit sphere.  Vertical dashed
line indicates the branch point $g_1^{\rm s} \approx 5.622$, at
which both coefficients diverge.}
\label{fig:sphere_C11}
\end{figure}

Figure \ref{fig:sphere_C11} shows the dependence of the coefficients
$C_{1,1}^{(g)}$ and $\Re\{C_{1,2}^{(g)}\}$ on $g$.  When $g$
approaches the branch point $g_1^{\rm s}$, the normalization of the
involved eigenfunctions $v_1^{(g)}$ and $v_2^{(g)}$ diverges,
resulting in the divergence of these coefficients: $C_{1,1}^{(g)} \to
+\infty$ and $\Re\{C_{1,2}^{(g)}\} \to -\infty$.  However, as
discussed in \cite{Moutal22}, these diverging contributions to the
signal compensate each other and thus imply no resonant behavior of
the signal near $g_1^{\rm s}$.  In other words, the signal changes
smoothly with $g$ even at the branch point.

\begin{figure}
\begin{center}
\includegraphics[width=75mm]{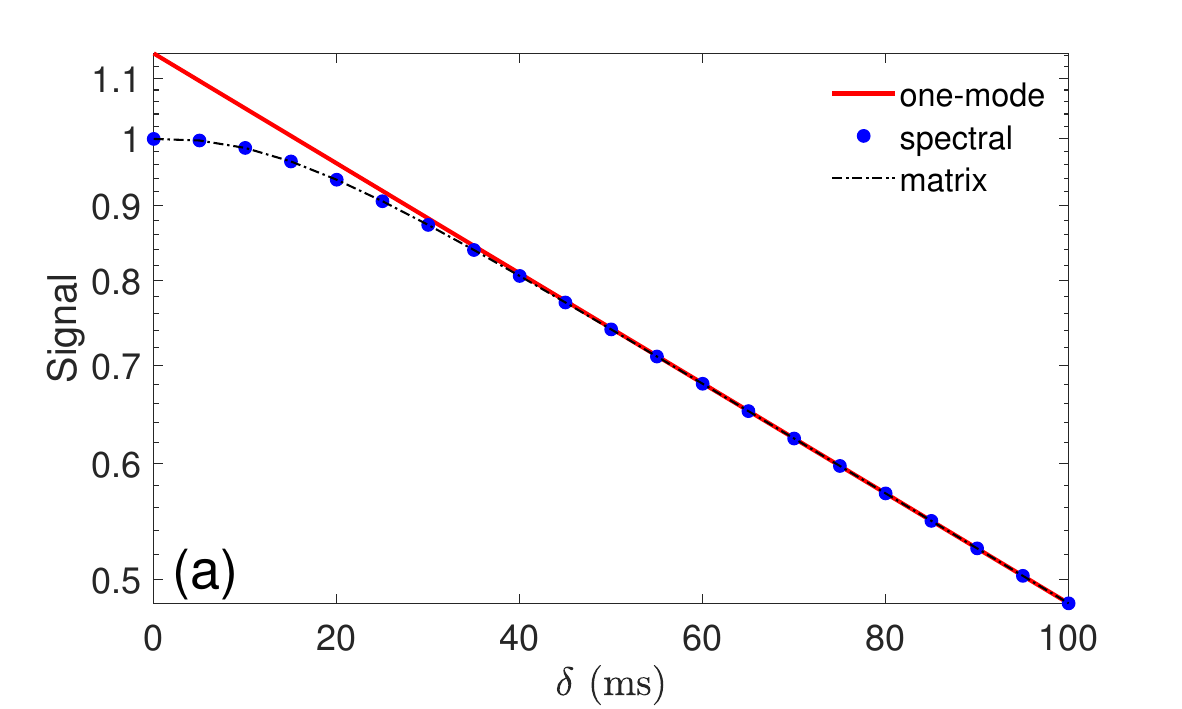} 
\includegraphics[width=75mm]{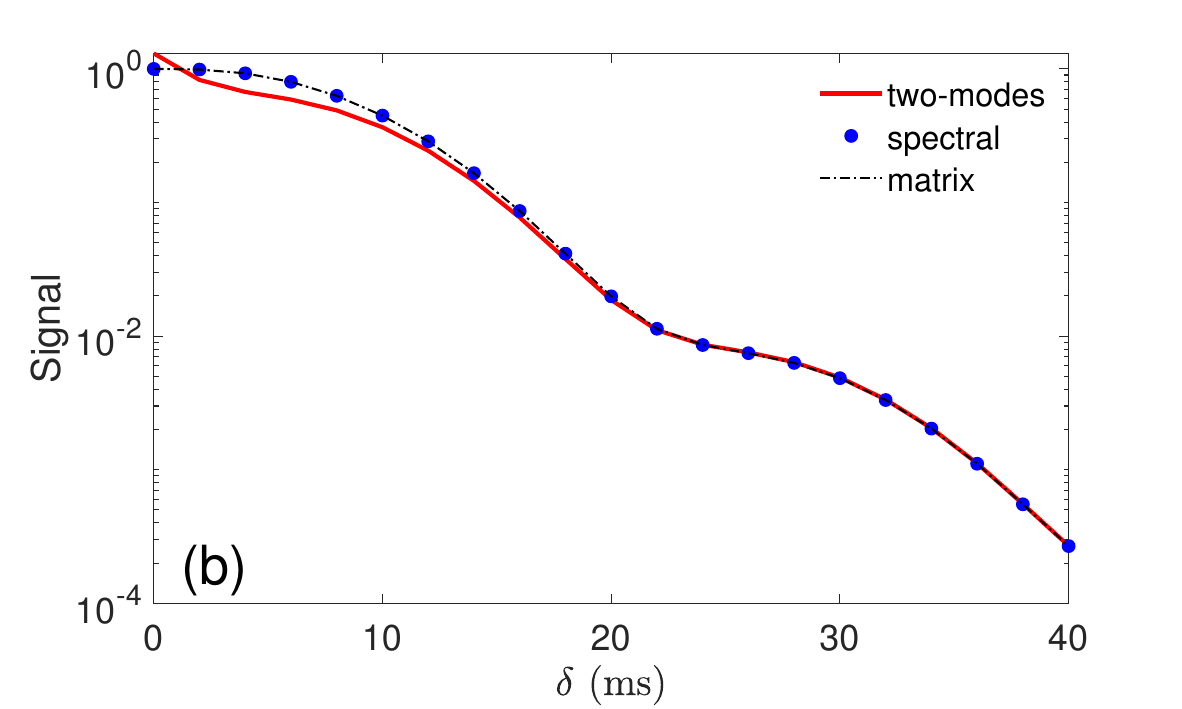} 
\end{center}
\caption{
Signal as a function of the gradient pulse duration $\delta$ for a
sphere of radius $R = 10~\mu$m, with $\gamma = 2.675\cdot
10^8$~rad/T/s (protons), $D_0 = 2.3\cdot 10^{-9}~\rm{m}^2/\rm{s}$
(water molecules), $G = 17$~mT/m {\bf (a)} and $G = 129$~mT/m {\bf
(b)}.  Circles present the exact spectral expansion
(\ref{eq:S_exact}), truncated to $333$ terms; dashed line indicates
the exact matrix representation (\ref{eq:S_matrix}) with the same
truncation; solid lines show one-mode and two-modes approximations
(\ref{eq:Sone}, \ref{eq:Stwo}) for panels (a) and (b), respectively.
Their parameters are: $R^2 \lambda_1^{(g)} \approx 0.188$ and $C_{1,1}
\approx 1.14$ for $G = 17$~mT/m (or $g = 2$, panel (a)); and $R^2
\lambda_1^{(g)} \approx 4.67 + 6.68i$, $C_{1,1} \approx 1.12$ and
$C_{1,2} \approx -0.46 + 0.18i$ for $G = 129$~mT/m (or $g = 15$, panel
(b)).}
\label{fig:sphere_signal}
\end{figure}

The accuracy of the approximations (\ref{eq:Sone}, \ref{eq:Stwo}) is
illustrated on Fig. \ref{fig:sphere_signal}.  Expectedly, both
approximations fail at very small $\delta$ when many eigenfunctions
are needed in the spectral expansion (\ref{eq:S_exact}) to get the
signal.  In turn, both approximations become accurate at larger
$\delta$.  On panel (b), one can also notice oscillations due to the
second term in Eq. (\ref{eq:Stwo}).  Their period is controlled by the
imaginary part of $\lambda_1^{(g)}$.  At high gradients, the leading
term of the large-$g$ asymptotic expansion of $\lambda_1^{(g)}$ is $g
R$ \cite{deSwiet94,Grebenkov18,Moutal19}, so that the last factor in
Eq. (\ref{eq:Stwo}) is approximately $e^{2i\gamma G R\delta}$.  One
thus retrieves a diffusion-diffraction pattern
\cite{Callaghan91,Cotts91,Sen95,Gibbs97,Ozarslan07}, which is more
common for short gradient pulses (see also a comparison between the
localization regime and narrow-pulse approximation in
\cite{Moutal20}).  Note also that the next-order corrections to
$\Im(\lambda_1^{(g)})$ can significantly alter this behavior.  In
turn, the real part of the first eigenvalue behaves at large $g$ as
\cite{deSwiet94,Moutal19}:
\begin{equation}  \label{eq:lambda1_asympt}
\Re\{\lambda_1^{(g)}\} = \frac{|a'_1|}{2 \ell_g^2} + \frac{1}{\sqrt{R} \ell_g^{3/2}} - \frac{\sqrt{3}}{4|a'_1| R \ell_g} 
+ O(\ell_g^{-1/2})  \qquad (g\to \infty),
\end{equation}
where $\ell_g = (\gamma G/D_0)^{-1/3} = g^{-1/3}$, and $a'_1 \approx
-1.02$ is the first zero of the derivative of the Airy function
${\mathrm{Ai}}(z)$.  Substituting the leading order of
Eq. (\ref{eq:lambda1_asympt}) into Eq. (\ref{eq:Stwo}), one retrieves
the stretched exponential decay of the signal: $\ln S \propto
G^{2/3}$.

\section{Bloch-Torrey operator for capped cylinders}
\label{sec:capped}

In this section, we consider restricted diffusion in a capped cylinder
of radius $R$ and height $H$: $\Omega =
\{ (x,y,z)\in\R^3 ~:~ x^2+y^2 < R^2, ~ -H/2 < z < H/2\}$.  Breaking
rotational invariance, this shape allows us to investigate how the
domain anisotropy can affect the spectrum of the Bloch-Torrey operator
for a gradient in an arbitrary direction.  Since the capped cylinder
is axisymmetric, there is no difference between $x$ and $y$ directions
so that one can focus on gradients in the $xz$ plane, for instance, by
setting $\G = G (\e_x \cos \eta + \e_z \sin\eta)$, where $\eta$ is the
angle with respect to the horizontal axis in the $xz$ plane.  The
corresponding Bloch-Torrey operator, denoted as $\B_g^{(\eta)}$, reads
in the cylindrical coordinates $(r,\theta,z)$ as
\begin{equation}  \label{eq:Beta_def}
\B_g^{(\eta)} = -\Delta + ig(x \cos\eta + z\sin\eta) = \B_{g\cos\eta}^{\rm d} + \B_{g\sin\eta}^{\rm i} ,
\end{equation}
where
\begin{equation}
\B_g^{\rm d} = -\biggl(\underbrace{\partial_r^2 + \frac{1}{r} \partial_r + \frac{1}{r^2} \partial_\theta^2}_{=\Delta_{\rm d}} 
\biggr) + ig r\cos\theta ,   \qquad   \B_g^{\rm i} = - \partial_z^2 + ig z
\end{equation}
are the Bloch-Torrey operators in the disk and in the interval,
respectively.  As these two operators act on different variables, the
eigenfunctions of $\B_g^{(\eta)}$ are factored, while its eigenvalues
are obtained as all possible sums of the eigenvalues of $\B_g^{\rm i}$
and $\B_g^{\rm d}$.  The operator $\B_g^{\rm i}$, also known as the
(complex) Airy operator \cite{Helffer}, was thoroughly studied in
\cite{Stoller91,Grebenkov14,Grebenkov17}, whereas $\B_g^{\rm d}$ was
analyzed in \cite{deSwiet94,Grebenkov18,Moutal19,Moutal22}.  We aim at
understanding how their spectral properties are superimposed in the
case of a capped cylinder.

The matrix elements of the operator $\B_g^{(\eta)}$ are derived in
\ref{sec:Acapped}.  In particular, the Laplacian eigenfunctions,
\begin{equation*}
u_{nklm}(r,\theta,z) \propto J_n(\alpha_{nk}r/R) s_l(n\theta) \cos(\pi m(z+H/2)/H),
\end{equation*}
are enumerated by multi-index $nkjm$, with $n = 0,1,2,\ldots$ being
the order of the Bessel function $J_n(z)$ of the first kind, $k =
0,1,2,\ldots$ being the index of the zeros $\alpha_{nk}$ of $J'_n(z)$,
$l$ distinguishing between $s_1(z) = \cos(z)$ and $s_2(z) = \sin(z)$,
and $m = 0,1,2,\ldots$ characterizing oscillations along $z$ axis.
The eigenvalues $\lambda_{nklm} = \alpha_{nk}^2/R^2 + \pi^2 m^2/H^2$
do not depend on $l$ and are in general either simple (for $n = 0$) or
twice degenerate (for $n > 0$), but higher degeneracies are possible.
As previously, we use the ordered sequence of these eigenvalues to
enumerate the eigenvalue branches $\lambda_j^{(g)}$ of the
Bloch-Torrey operators (see Table
\ref{tab:eigen_capped}).
Throughout this section, we fix $R = H = 1$ and then explore the
anisotropy by changing the gradient direction (angle $\eta$), as
explained below.

\begin{table}
\begin{center}
\begin{tabular}{c | c | c | c | c | c | c | c | c}
$j$               &  1   &  2-3     &  4-5     &  6   &  7-8     &  9    &  10-11    &  12-13   \\ \hline
$nklm$            & 0010 & 10(1-2)0 & 20(1-2)0 & 0011 & 10(1-2)1 & 0110  &  30(1-2)0 & 20(1-2)1 \\
$\lambda_j^{(0)}$ &  0   &  3.39    &  9.33    & 9.87 &  13.26   & 14.68 &   17.65   &  19.20   \\
\end{tabular}
\end{center}
\caption{
First 13 eigenvalues $\lambda_{nklm} = \lambda_j^{(0)}$ of the
(negative) Laplace operator, $\B_0 = -\Delta$, in the capped cylinder
with $R = H = 1$.  The position $j$ of the eigenvalue
$\lambda_j^{(0)}$ in the ordered sequence is used to enumerate the
branch $\lambda_j^{(g)}$ for $g \ne 0$.  Twice degenerate eigenvalues
(corresponding to $l = 1$ and $l=2$) are shown together, e.g.,
$\lambda_2^{(0)} = \lambda_3^{(0)}$. }
\label{tab:eigen_capped}
\end{table}

\subsection{Parallel and perpendicular directions of the gradient}

We start with two simple cases when the gradient is either aligned
with the cylinder axis and thus $\B_g^{(\pi/2)} = \B_g^z = \B_0^{\rm
d} + \B_g^{\rm i}$, or lies in the transverse $xy$ plane so that
$\B_g^{(0)} = \B_g^x = \B_g^{\rm d} + \B_0^{\rm i}$.

\begin{figure}
\begin{center}
\includegraphics[width=100mm]{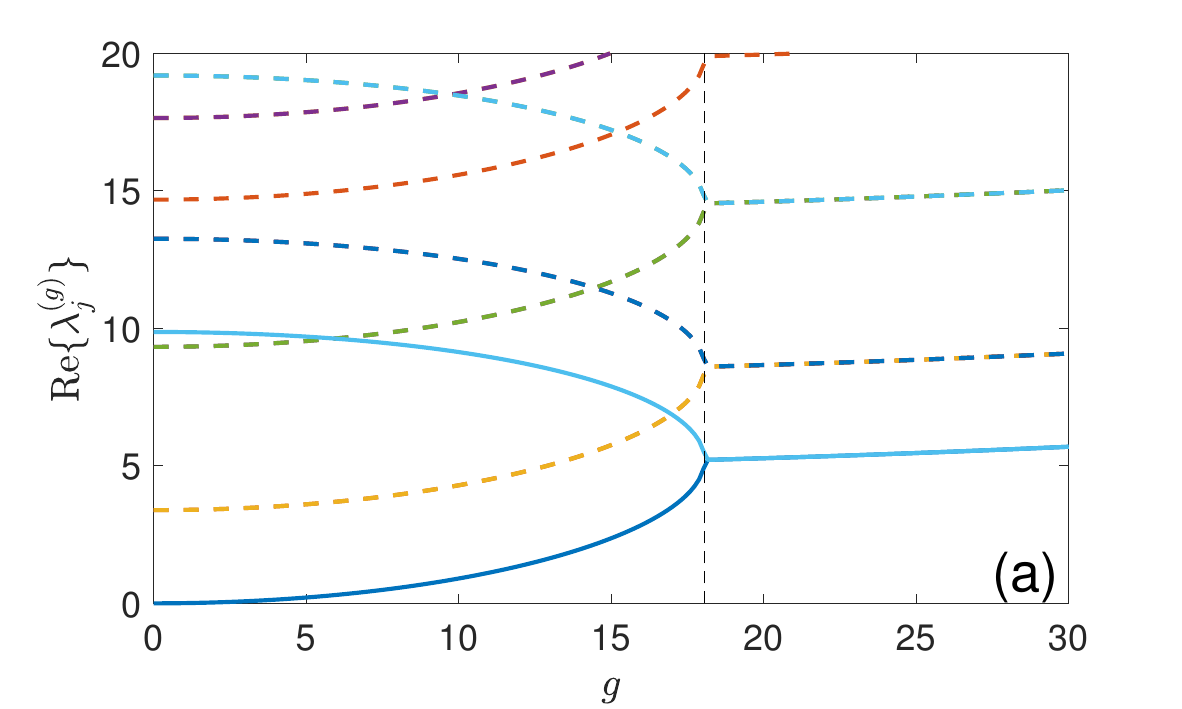} 
\includegraphics[width=100mm]{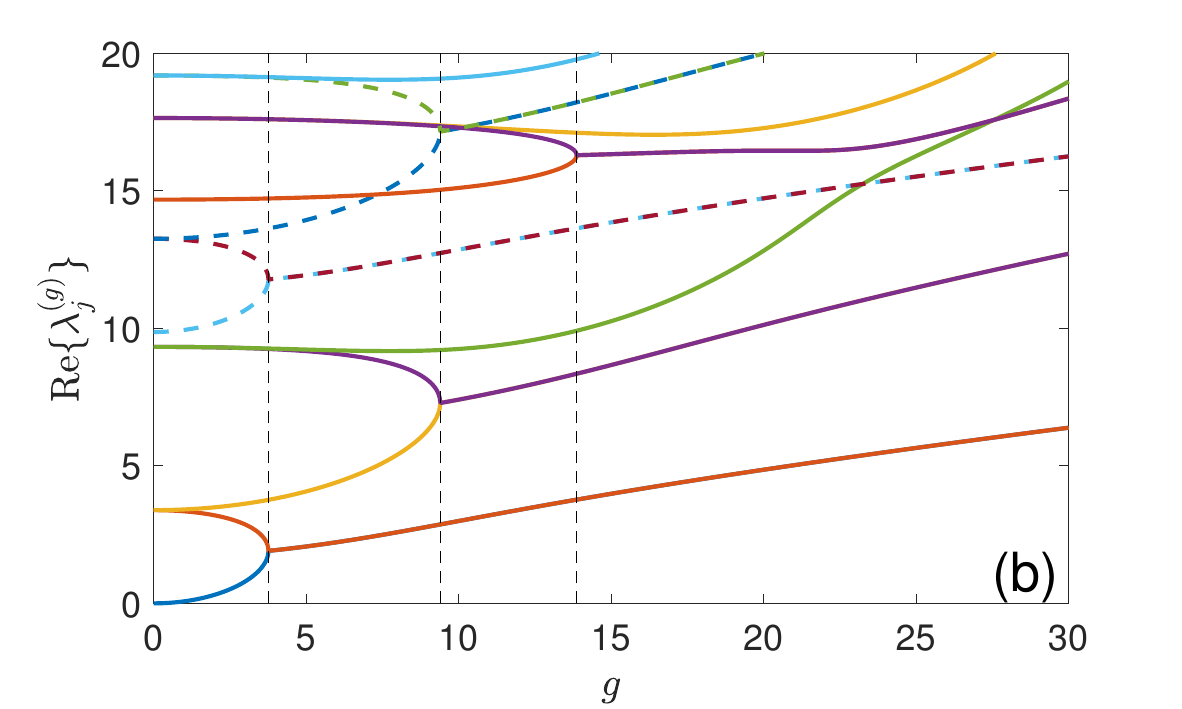} 
\end{center}
\caption{
Real part of the first 13 eigenvalues of the Bloch-Torrey operators
$\B_g^z$ {\bf (a)} and $\B_g^x$ {\bf (b)} for a capped cylinder with
$R = H = 1$.  Dashed lines indicate the eigenvalues that do not
contribute to the signal.  Vertical lines indicate the positions of
branch points: $g_1^{\rm i} \approx 18.06$ (a) and $g_1^{\rm d}
\approx 3.76$, $g_2^{\rm d} \approx 9.39$, $g_3^{\rm d} \approx 13.87$
(b).  Note that each pair of twice degenerate eigenvalues appears as a
single branch on panel (a).  Truncation order was $320$.  }
\label{fig:capped_spectrum}
\end{figure}

Figure \ref{fig:capped_spectrum} presents real parts of the first 13
eigenvalues of the Bloch-Torrey operators $\B_g^z$ (panel (a)) and
$\B_g^x$ (panel (b)).  Let us first inspect the spectrum of $\B_g^z$,
which is the sum of the (negative) Laplace operator $\B_0^{\rm d} =
-\Delta_{\rm d}$ in the disk and the Bloch-Torrey operator $\B_g^{\rm
i}$ on the interval $(-H/2,H/2)$.  The spectrum of $\B_g^{\rm i}$ was
thoroughly investigated in \cite{Stoller91,Grebenkov14,Grebenkov17};
for instance, Fig. 3 from \cite{Stoller91} shows the real and
imaginary parts of several eigenvalues.  In particular, the branch
of the first two eigenvalues $\lambda_1^{(g)}$ and $\lambda_2^{(g)}$
can be retrieved in Fig. \ref{fig:capped_spectrum}(a), which zooms out
Fig. 3 from \cite{Stoller91} to a smaller range of $g$.  Moreover,
Stoller {\it et al.}  studied the branch points of $\B_g^{\rm i}$
and found an explicit formula \cite{Stoller91}, which reads in our
notations (see also discussion in \cite{Moutal22}):
\begin{equation}
g_{k}^{\rm i} = \sqrt{3} \frac{27}{4} j_k^2 \,,    \qquad \textrm{where} \quad J_{-2/3}(j_k) = 0 \quad (k=1,2,\ldots).
\end{equation}
In particular, one gets $g_1^{\rm i} \approx 18.06$ and $g_2^{\rm i}
\approx 229.35$.  The position of the first branch point $g_1^{\rm i}$,
as indicated by the vertical line, is in excellent agreement with this
prediction.
The major difference between Fig. \ref{fig:capped_spectrum}(a) and
Fig. 3 from \cite{Stoller91} is that the pair of eigenvalues
$\lambda_1^{(g)}$ and $\lambda_2^{(g)}$ is replicated and shifted
vertically by adding the eigenvalues of $-\Delta_{\rm d}$, e.g.,
$3.3900$, $9.3284$, $14.6820$, etc.  As a consequence, there are
infinitely many pairs of eigenvalues that branch at each value
$g_1^{\rm i}, g_2^{\rm i}, \ldots$.  Note that the ``shifted''
eigenvalues correspond to the eigenfunctions $u_{nkl}^{\rm d}$ of
$-\Delta_{\rm d}$ that are orthogonal to $u_{001}^{\rm d} = const$, so
that the resulting eigenfunctions $v_j^{(g)}$ do not contribute to the
signal.

Let us now look at the spectrum of the Bloch-Torrey operator $\B_g^x$,
which is the sum of $\B_g^{\rm d}$ and the second derivative
$\B_0^{\rm i} = -\partial_z^2$ on the interval.  The spectrum of the
former operator is simply replicated and shifted vertically by the
eigenvalues $\pi^2 m^2/H^2$ ($m=1,2,\ldots$) of $-\partial_z^2$.
These shifted eigenvalues are shown by dashed lines in
Fig. \ref{fig:capped_spectrum}(b) because the associated
eigenfunctions do not contribute to the signal due to the presence of
the factor $\cos(\pi m(z+H/2)/H)$, whose integral over the interval
$(-H/2,H/2)$ vanishes for any $m > 0$.  In turn, the eigenvalues shown
by solid lines correspond to $m = 0$ and repeat the spectrum of the
Bloch-Torrey operator in the disk (compare with Fig. 9 from
\cite{Moutal22}).  In particular, one observes here three branch
points at $g_1^{\rm d} \approx 3.76$, $g_2^{\rm d} \approx 9.39$, and
$g_3^{\rm d} \approx 13.87$, which are replicated along the vertical
axis by adding $\pi^2 m^2/H^2$.

\subsection{Changing gradient direction}

The structure of the spectra for both considered operators $\B_g^z$
and $\B_g^x$ was rather simple because one of two terms in
Eq. (\ref{eq:Beta_def}) was independent of $g$ and thus just shifted
vertically the spectrum of the other.  For intermediate angles $\eta$,
both terms in Eq. (\ref{eq:Beta_def}) depend on $g$, and the angle
$\eta$ controls rescaling of each spectrum through the factors
$g\cos\eta$ and $g\sin\eta$.  Changing $\eta$, one can ``tune''
continuously the spectra of $\B_{g\cos\eta}^{\rm d}$ and
$\B_{g\sin\eta}^{\rm i}$, and see how their features change.  Most
importantly, even though the eigenfunctions are still factored along
the longitudinal and transverse directions $\e_z$ and $\e_x$, the
nonzero gradients along $\e_x$ and $\e_z$ directions break the
symmetries of these factors so that all eigenfunctions may contribute
to the signal.

\begin{figure}
\begin{center}
\includegraphics[width=100mm]{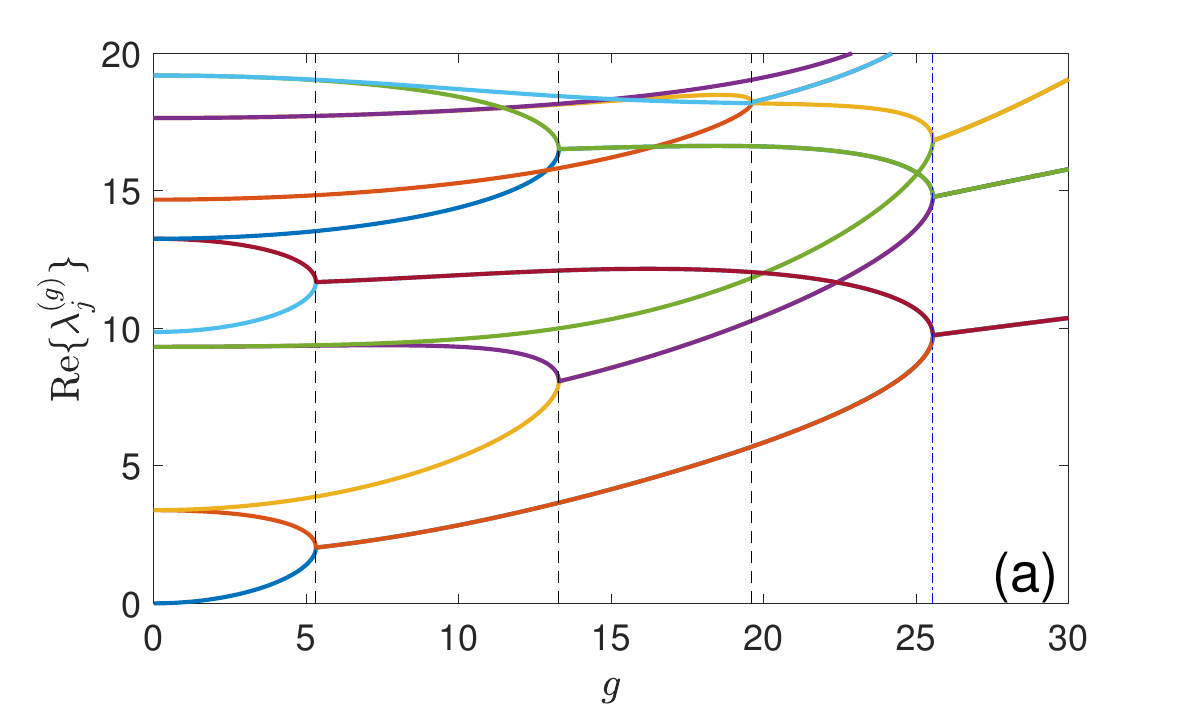} 
\includegraphics[width=100mm]{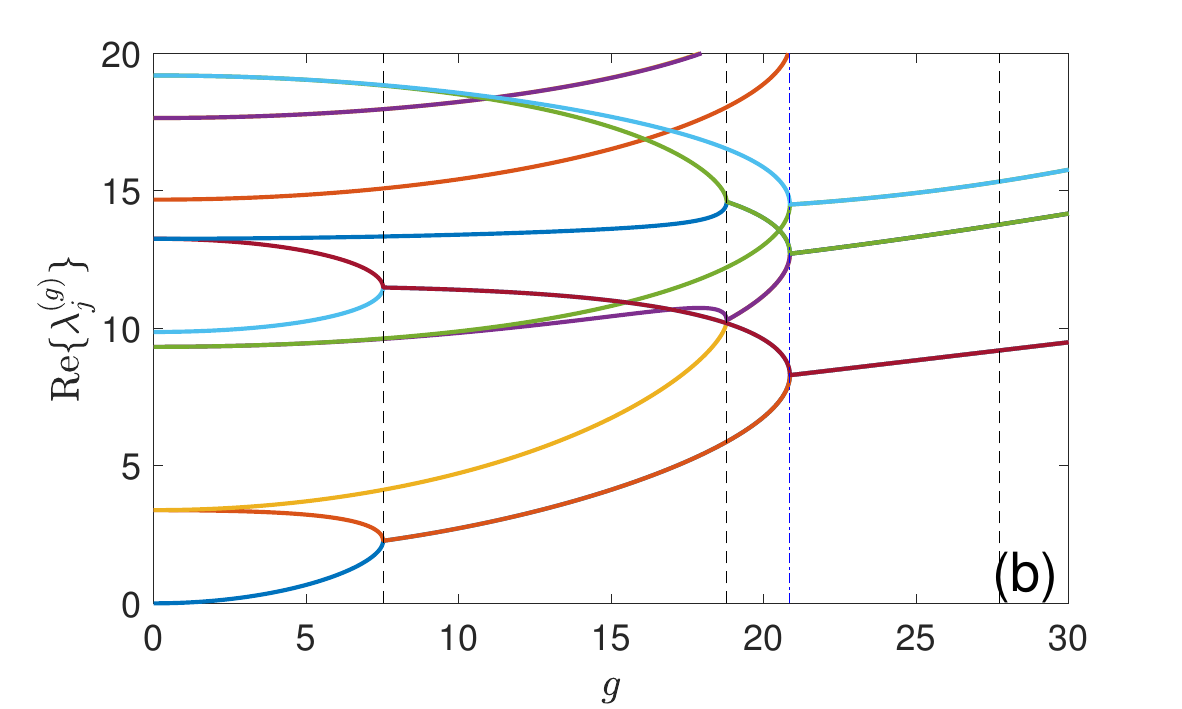} 
\end{center}
\caption{
Real part of the first 13 eigenvalues of the Bloch-Torrey operator
$\B_g^{(\eta)}$ for a capped cylinder with $R = H = 1$, and the
gradient applied in the $xz$ plane at the angle $\eta$ with respect to
the $x$-axis: $\eta = \pi/4$ {\bf (a)} and $\eta = \pi/3$ {\bf (b)}.
Vertical dashed lines show the positions of rescaled branch points
$g_1^{\rm d} \approx 3.76/\cos(\eta)$, $g_2^{\rm d} \approx
9.36/\cos(\eta)$, $g_3^{\rm d} \approx 13.87/\cos(\eta)$, and
$g_1^{\rm i} \approx 18.06/\sin(\eta)$ associated to the operators
$\B_g^{\rm d}$ and $\B_g^{\rm i}$.  Note that each pair of twice
degenerate eigenvalues appears as a single branch on panel (b).
Truncation order was $320$.  }
\label{fig:capped_spectrum2}
\end{figure}

Figure \ref{fig:capped_spectrum2} shows the real parts of the first
eigenvalues of the Bloch-Torrey operator $\B_g^{(\eta)}$ for two
angles: $\eta = \pi/4$ and $\eta = \pi/3$.  One sees how the two
spectra of $\B_{g\cos\eta}^{\rm d}$ and $\B_{g\sin\eta}^{\rm i}$ are
superimposed and tuned by the angle.  In particular, as $g$ is
multiplied by $\cos\eta$ or $\sin\eta$, the branch points of the
spectra are rescaled by $1/\cos\eta$ and $1/\sin\eta$, respectively.
Changing the gradient angle, one can continuously shift the positions
of branch points and thus re-organize the branch structure of
the spectrum.

\begin{figure}
\begin{center}
\includegraphics[width=120mm]{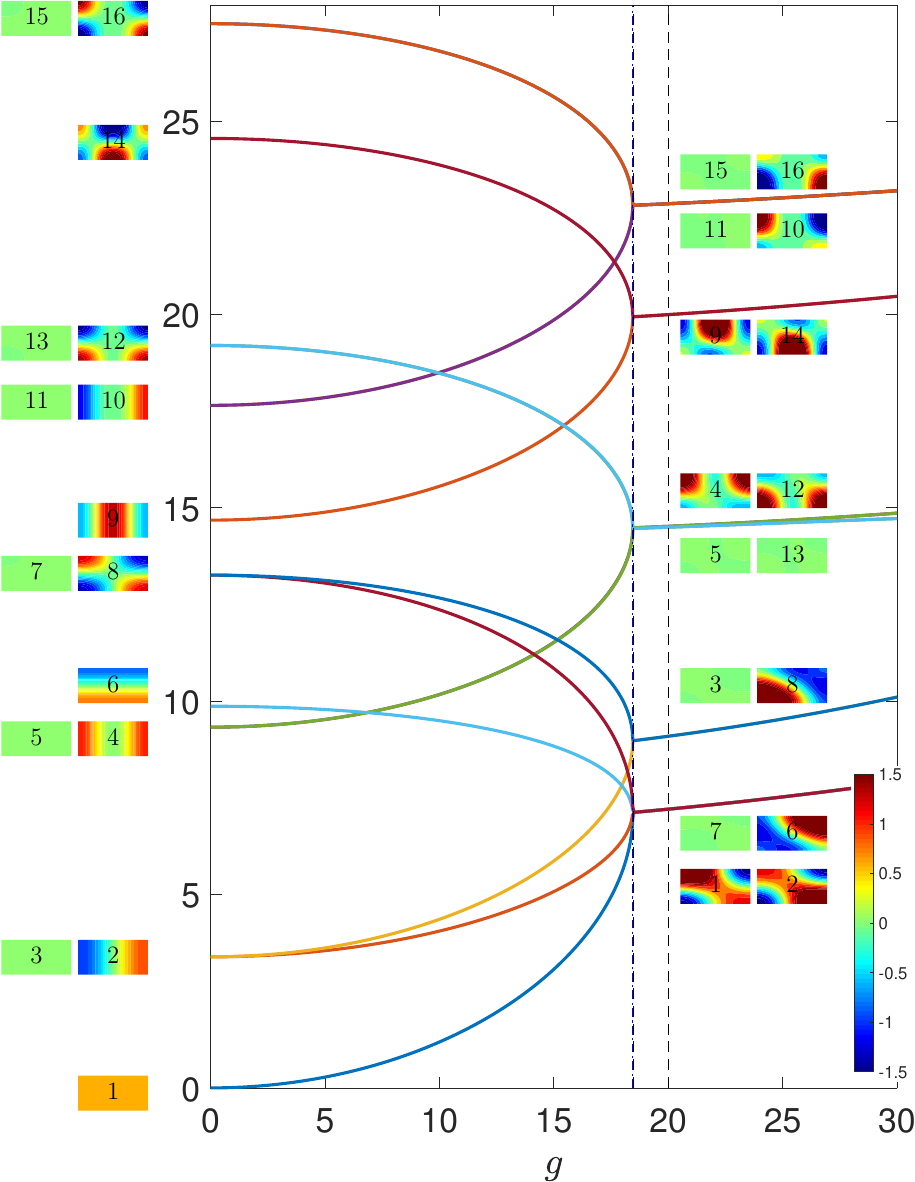} 
\end{center}
\caption{
Real part of the first 16 eigenvalues $\lambda_j^{(g)}$ of the
Bloch-Torrey operator $\B_g^{(\eta)}$ for the capped cylinder with $R
= H = 1$, and the gradient applied in the $xz$ plane at the angle
$\eta = \tan^{-1}(18.06/3.76) \approx 1.3661 \approx 78.3^\circ$ with
respect to $x$ axis.
Colored snapshots show the $xz$ projection of the real part of the
corresponding eigenfunction, evaluated at $g = 0$ (on the left) and at
$g = 20$ (near vertical dashed line).  Color indicates changes of
$\Re\{v_j^{(g)}\}$ from $-1.5$ (dark blue) to $1.5$ (dark red), with
the colorbar shown at right bottom, being the same for all snapshots.
The branch point is located at $18.5$.  The eigenvalues and
eigenfunctions were constructed via the matrix formalism (see
\ref{sec:numerics}), in which the matrices were truncated at 320.  }
\label{fig:capped_spectrum3}
\end{figure}

Let us now focus on the particular setting, shown in
Fig. \ref{fig:capped_spectrum3}, in which the angle $\eta$ is tuned to
make equal the rescaled branch points $g_1^{\rm d}/\cos\eta$ and
$g_1^{\rm i}/\sin\eta$, i.e., $\eta =
\tan^{-1}(g_1^{\rm d}/g_1^{\rm i}) \approx 1.3661 \approx 78.3^\circ$.
The position $g_1^{\rm c} = g_1^{\rm d}/\cos\eta \approx 18.5$ of the
first group of branch points is indicated by a vertical line.  As
compared to Fig. \ref{fig:capped_spectrum2}, there are more branches
that merge at a single branch point.  For instance, there are four
eigenvalues $\lambda_1^{(g)}, \lambda_2^{(g)}, \lambda_6^{(g)},
\lambda_7^{(g)}$ that merge at the first branch point, two eigenvalues
$\lambda_3^{(g)}$ and $\lambda_8^{(g)}$ that merge at the second one,
and four eigenvalues $\lambda_4^{(g)}, \lambda_5^{(g)},
\lambda_{12}^{(g)}, \lambda_{13}^{(g)}$ that merge at the third one.
Curiously, these last four eigenvalues form two {\it distinct}
branches for $g > g_1^{\rm c}$: one pair $\lambda_4^{(g)},
\lambda_{12}^{(g)}$ and the other pair $\lambda_5^{(g)},
\lambda_{13}^{(g)}$ (one can notice a small deviation between them at
$g = 30$, which is further increased at larger $g$).  It turns out
that the eigenvalues in each group keep the same value of the index
$l$ that distinguished $s_1(n\theta) = \cos(n\theta)$ and
$s_2(n\theta) = \sin(n\theta)$ in the eigenfunctions of the Laplace
operator.  According to Table \ref{tab:eigen_capped}, one sees that
the eigenvalues with indices $j = 1,2,6,7$ have $l = 1$, those with $j
= 3,8$ have $l = 2$, those with $j = 4,12$ have $l=1$, and those with
$j = 5,13$ have $l = 2$, etc.  This is expected because the gradient
is applied in the $xz$ plane and thus preserves the distinction
between $l = 1$ and $l = 2$ in the angular dependence (in the similar
way as the gradient applied along $z$ axis preserved the dependence on
the angle $\phi$ for a sphere, as discussed above).
The insets of Fig. \ref{fig:capped_spectrum3} illustrate the real
parts of the $xz$ projections of the associated eigenfunctions
$v_j^{(g)}$ at $g = 0$ and $g = 20$.  One sees how the symmetries of
eigenfunctions drastically change at the branch point.

\begin{figure}
\begin{center}
\includegraphics[width=\textwidth]{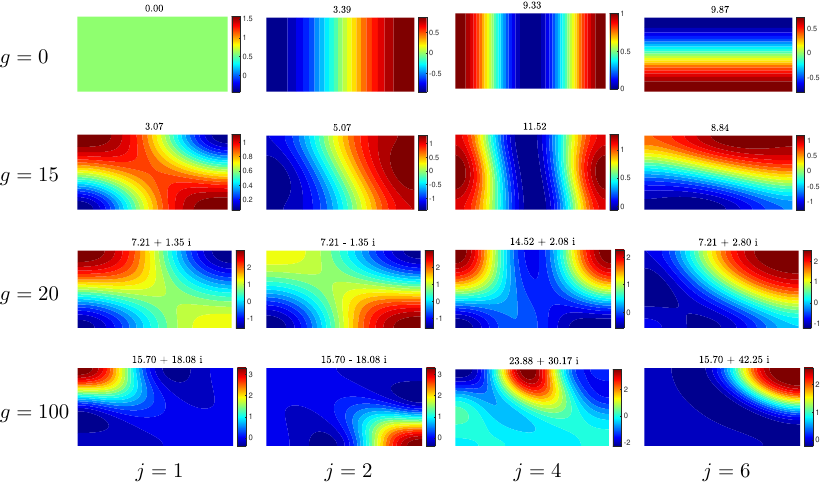} 
\end{center}
\caption{
$xz$ projection of the real part of four eigenfunctions $v_j^{(g)}$ of
the Bloch-Torrey operator $\B_g^{(\eta)}$ for a capped cylinder with
$R = H = 1$.  Different columns correspond to $j\in \{1,2,4,6\}$,
while different rows correspond to $g \in \{0, 15, 20, 100\}$.  The
gradient is applied in the $xz$ plane with the angle $\eta =
\tan^{-1}(18.06/3.76) \approx 1.3661 \approx 78.3^\circ$ with respect
to the $x$-axis.  The associated eigenvalue is indicated on the top of
each plot.  The branch point is located at $18.5$.  Truncation
order was 320. }
\label{fig:capped_eigenmodes}
\end{figure}

This is further illustrated on Fig. \ref{fig:capped_eigenmodes},
showing the real part of the $xz$ projection of four eigenfunctions
$v_1^{(g)}$, $v_2^{(g)}$, $v_4^{(g)}$ and $v_6^{(g)}$ at different
$g$.  As earlier for the case of a sphere, the constant eigenfunction
$v_1^{(0)}$ is rapidly destroyed by the gradient; in turn, the
symmetries of other three eigenfunctions $v_2^{(0)}$, $v_4^{(0)}$ and
$v_6^{(0)}$ are still visible (though slightly perturbed) at $g = 15$,
which is below the branch point $18.5$.  When $g$ exceeds the
branch point, the symmetries change, and the eigenfunctions start
to be more and more localized.  At first thought, the observed
localization pattern is puzzling.  In fact, for a smooth boundary, the
localization occurs at specific boundary points $\x_b$, at which the
normal vector $\n(\x_b)$ is parallel to the gradient $\G$
\cite{Grebenkov18}.  In other words, the gradient direction determines
the location of the localized eigenfunctions on the boundary.  As the
boundary of the capped cylinder is not smooth, the asymptotic analysis
from \cite{Grebenkov18} is not applicable.  Moreover,
Fig. \ref{fig:capped_eigenmodes} shows that the above selection rule
is actually not valid here.  In fact, the gradient is directed at the
angle $\eta \approx 78.3^\circ$ in the $xz$ plane with respect to the
$x$ axis.  One might thus expect localization at left bottom and right
top corners.  However, the eigenfunctions $v_1^{(g)}$ and $v_2^{(g)}$
at $g = 100$ are localized at the other corners, namely, the left up
and the right bottom corners, respectively.  Moreover, $v_4^{(g)}$ is
localized in the middle of the upper edge.  For a capped cylinder,
this behavior can be explained by the factored structure of
eigenfunctions.  As discussed earlier, every eigenfunction $v_j^{(g)}$
is the product of an eigenfunction of $\B_g^{\rm d}$ and an
eigenfunction of $\B_g^{\rm i}$ for the disk and the interval,
respectively.  At high enough $g$, both factors are localized: the
eigenfunction of $\B_g^{\rm d}$ is localized at either of two opposite
points of the disk along the $x$ axis, while the eigenfunction of
$\B_g^{\rm i}$ is localized at either of two endpoints of the interval
$(-H/2,H/2)$ along the $z$ axis.  The eigenfunction $v_j^{(g)}$ can
thus be localized in or near any corner of the $xz$ projection.
Further investigation of the localization in domains with nonsmooth
boundaries presents an open mathematical problem.

\subsection{Exploring the anisotropy}

The above illustrations of the spectral properties of the Bloch-Torrey
operator were realized for a particular capped cylinder with $H = R =
1$.  While it is easy to replicate the above results for any capped
cylinder, the overall structure of the spectrum does not change
significantly.  In fact, one can rewrite the Bloch-Torrey operator in
the capped cylinder of radius $R$ and height $H$ as
\begin{equation}
\B_g = \frac{1}{R^2} \bigl(-\bar{\Delta}_{\rm d} + i g\cos\eta \, R^3 \, \bar{r} \cos\theta\bigr)
+ \frac{1}{H^2} \bigl(-\partial^2_{\bar{z}} + ig\sin\eta \, H^3 \, \bar{z}\bigr) ,
\end{equation}
where bar denotes rescaled quantities: $\bar{r} = r/R$, $\bar{z} =
z/H$, $\bar{\Delta}_{\rm d} = R^2\Delta_{\rm d}$.  Setting $\bar{g} =
g \sqrt{R^6 \cos^2\eta + H^6 \sin^2\eta}$ and introducing the angle
$\bar{\eta}$ such that $\tan(\bar{\eta}) = (H^3/R^3) \tan(\eta)$, one
has
\begin{equation}  \label{eq:B_sum}
\B_g = \frac{1}{R^2} \bar{B}_{\bar{g}\cos\bar{\eta}}^{\rm d} + \frac{1}{H^2} \bar{B}_{\bar{g}\sin\bar{\eta}}^{\rm i} \,,
\end{equation}
where the first term is the Bloch-Torrey operator in the unit disk
(divided by $R^2$) with the dimensionless gradient $\bar{g}
\cos\bar{\eta}$, whereas the second term is the Bloch-Torrey operator in
the unit interval (divided by $H^2$) with the dimensionless gradient
$\bar{g} \sin\bar{\eta}$.  In the previous subsection, we used $R = H
= 1$ and thus considered the sum of these two basic operators.  In
general, the prescribed lengths $R$ and $H$ re-define the gradient
amplitude (from $g$ to $\bar{g}$) and the gradient angle (from $\eta$
to $\bar{\eta}$), as well as weighting factors $1/R^2$ and $1/H^2$ in
the linear combination (\ref{eq:B_sum}).  The structure of the
eigenfunctions of $\B_g$ is thus affected by anisotropy of the capped
cylinder only through $\bar{g}$ and $\bar{\eta}$; in turn, the
spectrum is also controlled by the weights $1/R^2$ and $1/H^2$ that
can reshape eigenvalue branches and shift the branch points.
Developing experimental protocols that are sensitive to the shape of
eigenfunctions will potentially allow to exploit this property in
order to probe microscopic anisotropy of porous media at high
gradients.

\section{Conclusion}
\label{sec:conclusion}

In this paper, we investigated the spectral properties of the
Bloch-Torrey operator $\B_g = -\Delta + igx$ in two three-dimensional
domains: a sphere and a capped cylinder.  These shapes are typical
models in diffusion MRI, representing, e.g., a soma and a neuron in
the brain tissue.  While the general asymptotic behavior of
eigenvalues and eigenfunctions was known in the limits of small and
large $g$, there is no spectral result for the intermediate range of
$g$, which is the most relevant for applications.  In particular, we
studied the structure of the spectrum, the dependence of eigenvalue
branches on $g$, the existence of branch points, and drastic symmetry
changes of eigenfunctions at branch points.  Despite the geometric
simplicity of the considered domains, we had to rely on the matrix
formalism to construct the eigenvalues and eigenfunctions of $\B_g$
numerically.  We illustrated how different eigenvalue branches merge
at branch points and how the symmetries of eigenfunctions, inherited
from the Laplace operator, are destroyed at these points.  For a
capped cylinder, we also showed the effect of anisotropy on the
spectrum, in particular, how rotating the gradient direction allows
one to rescale the spectra of the operators $\B_g^{\rm d}$ and
$\B_g^{\rm i}$ in the orthogonal directions and thus to tune the
branch points.  The localization of eigenfunctions was shown to occur
near the points with $z = \pm H/2$ and $r = R$, at which the
cylindrical wall joins the top and bottom caps.  As the boundary of a
capped cylinder is not smooth at these points, the asymptotic behavior
established in \cite{Grebenkov18} is not applicable, and further
analysis of localization in domains with nonsmooth boundaries is
needed.

The present work lays the theoretical ground for various applications
in diffusion MRI.  As shown earlier, the macroscopic signal is getting
more sensitive to the microstructure at high gradients
\cite{Grebenkov14,Grebenkov18b}.  As a consequence, high-gradient
diffusion MRI is a promising research direction with potential
applications in material sciences, neurosciences and medicine
\cite{Wedeen12,Huang21,Williamson23}.  An intuitive explanation of
this enhanced sensitivity is that the localized eigenfunctions may
probe selected boundary regions and thus access refined information on
the microstructure.  For instance, the coefficients $C_{j,j'}^{(g)}$
from Eq. (\ref{eq:C_def}) of the spectral expansion (\ref{eq:S_exact})
are sensitive to the overlap between two eigenfunctions and thus may,
potentially, probe a sort of spatial correlations between different
boundary regions, at which these eigenfunctions are localized.
Moreover, the access to the Bloch-Torrey operators $\B_g^x$, $\B_g^y$
and $\B_g^z$ for three orthogonal directions allows one to analyze
double-pulsed field-gradient experiments
\cite{Callaghan02,Komlosh07,Ozarslan08,Ozarslan09,Ozarslan09b,Jespersen13}
in terms of spectral expansions similar to Eq. (\ref{eq:S_exact}).  In
particular, as the coefficients of such expansions involve different
eigenfunctions of these non-commuting Bloch-Torrey operators, one may
potentially reveal additional information on the microstructure such
as its local anisotropy or curvature.  More generally, an experimental
exploration of eigenfunctions symmetry changes at branch points
presents a very interesting but challenging task, and the developed
spectral approach may pave a way towards new imaging modalities at
high gradients.

\section*{Data availability statement}

No new data were created or analysed in this study.

\section*{Acknowledgments}

The author acknowledges the Alexander von Humboldt Foundation for
support within a Bessel Prize award.

\appendix
\section{Computation of spectral properties}
\label{sec:numerics}

In this Appendix, we extend the description of the numerical procedure
from \cite{Moutal22} that we use for computing the eigenvalues and
eigenfunctions of the Bloch-Torrey operator $\B_g$.  It is inspired
from the matrix formalism \cite{Barzykin98,Grebenkov07,Grebenkov08},
in which the magnetization is decomposed onto the complete basis of
Laplacian eigenfunctions $u_k$ with Neumann boundary condition, which
are known explicitly for simple domains (e.g., a disk and a sphere).
Throughout this Appendix, we do not discuss mathematical aspects of
the problem such the convergence of spectral representations.  Our
goal here is to provide a practical recipe for numerical computations.
In order to deal with dimensionless quantities, we will rescale
lengths by the ``size'' $R$ of the confining domain $\Omega$, e.g., by
its (half-)diameter (this choice does not matter in practice).  For
the examples considered in the paper, $R$ is the radius of the sphere
or of the capped cylinder.

We search an eigenfunction $v_j^{(g)}$ of the Bloch-Torrey operator as
\begin{equation}  \label{eq:vn_uk}
v_j^{(g)}(\x) = \sum\limits_k X_{j,k}^{(g)} u_k(\x),
\end{equation}
with unknown coefficients $X_{j,k}^{(g)}$.  Substituting
Eq. (\ref{eq:vn_uk}) into the eigenvalue problem (\ref{eq:eigen_def}),
one gets
\begin{equation*}
\fl
\lambda_j^{(g)} \sum\limits_k X_{j,k}^{(g)} u_k(\x) = \lambda_j^{(g)} v_j^{(g)}(\x) = (-\Delta + ig x)v_j^{(g)}(\x) = 
\sum\limits_k X_{j,k}^{(g)} (-\Delta + ig x) u_k(\x).
\end{equation*}
Multiplying this equation by $u_{j'}^*(\x) R^2$, integrating over
$\Omega$, and using orthogonality of Laplacian eigenfunctions, we get
for any $j'$
\begin{equation}
R^2 \lambda_j^{(g)} X_{j,j'}^{(g)} = \sum\limits_k X_{j,k}^{(g)} (\Lambda_{k,j'} + i\bar{g} B_{k,j'}) ,
\end{equation}
where $\Lambda_{k,j'} = \delta_{k,j'} \lambda_k R^2$, $B_{k,j'} =
\int\nolimits_\Omega d\x\, u_k(\x) \, (x/R) \, u^*_{j'}(\x)$,
$\lambda_k$ are the eigenvalues of the (negative) Laplace operator
$-\Delta$, and we introduced the dimensionless parameter $\bar{g} =
gR^3 = R^3 \gamma G/D_0$.  The multiplication by $R^2$ ensured that
both matrices $\Lambda$ and $B$ are dimensionless.  In a matrix form,
one has
\begin{equation}  \label{eq:LambdaX}
\Lambda^{(g)} X = X (\Lambda + i \bar{g} B),
\end{equation}
where $\Lambda^{(g)}$ is the diagonal matrix of eigenvalues $R^2
\lambda_n^{(g)}$ of the Bloch-Torrey operator $\B_g$.  
As a consequence, the diagonalization of the matrix $\Lambda + i
\bar{g} B$ yields $\Lambda^{(g)}$ and $\tilde{X}$, where $\tilde{X}$
is the matrix whose columns contain {\it left eigenvectors}, from
which $X$ is obtained by complex-conjugate transpose: $X =
\tilde{X}^{\dagger,*}$, where $\dagger$ denotes transpose without
complex conjugation, i.e., $[X^\dagger]_{k,j} = X_{j,k}$.  For
instance, one could use the matlab commands
\begin{center}
\verb|[V,LambdaG,Xtilde] = eig(Lambda + 1i*gbar*B);   X = Xtilde'; |
\end{center}
to get $\Lambda^{(g)}$ and $X$.

To ensure the normalization (\ref{eq:vj_norm}) of eigenfunctions, one
can use the representation (\ref{eq:vn_uk}) that implies
\begin{equation}  \label{eq:vj_vj_XUX}
\int\limits_{\Omega} d\x \, v_j^{(g)}(\x) \, v_{j'}^{(g)}(\x)  = [X W X^\dagger]_{j,j'} = \delta_{j,j'} ,
\end{equation}
where
\begin{equation}  \label{eq:U_def}
W_{k,k'} = \int\limits_{\Omega} d\x \, u_k(\x) \, u_{k'}(\x).
\end{equation}
The eigenfunctions $\{u_k\}$ of the Laplace operator with
Neumann boundary condition can be chosen to be real-valued, in which
case $W$ is the identity matrix.  However, it may also be convenient
to employ complex-valued Laplacian eigenfunctions.  In this more
general setting, the scalar product in $L_2(\Omega)$ includes
complex-conjugation; as a consequence, even though two eigenfunctions
$u_k$ and $u_{k'}$ are orthogonal to each other,
$(u_k,u_{k'})_{L_2(\Omega)} = \int\nolimits_{\Omega} u_{k}^* u_{k'} =
0$, their integral in Eq. (\ref{eq:U_def}) may not be zero, and the
matrix $W$ is not necessarily identity.  For instance, we used this
convention in Sec. \ref{sec:sphere} by choosing $u_{nkm}
\propto e^{im\phi}$ that yields some nonzero off-diagonal elements of
the matrix $W$ in Eq. (\ref{eq:U_sphere}) for a sphere.  We stress
that this purely conventional issue does not affect any spectral
property of the Bloch-Torrey operator.

We also recall that the integral in Eq. (\ref{eq:vj_vj_XUX}) may be
zero for $j = j'$ at specific values of $g$.  Moreover, if an
eigenvalue $\lambda_j^{(g)}$ is degenerate, the associated
eigenfunctions form an eigenspace so that the coefficients
$X_{j,k}^{(g)}$ are not defined uniquely but up to a rotation in that
eigenspace (see \ref{sec:A_ortho} for a simple orthogonalization
procedure).  We stress that this ambiguity does not affect the
resulting macroscopic signal but may render the interpretation of
eigenfunctions more sophisticated.

According to Eq. (\ref{eq:C_def}), the coefficients $C_{j,j'}^{(g)}$
from the spectral expansion (\ref{eq:S_exact}) of the signal can be
written as
\begin{equation}
C_{j,j'}^{(g)} = \mu_j^{(-g)} \, \Gamma_{j,j'}^{(g)} \, \mu_{j'}^{(g)} ,
\end{equation}
where
\begin{equation}  \label{eq:mu_def}
\mu_j^{(g)} = \frac{1}{\sqrt{|\Omega|}} \int\limits_\Omega d\x \, v_j^{(g)}(\x)
\end{equation}
is the projection of the eigenfunction $v_j^{(g)}(\x)$ onto a
constant, and 
\begin{equation}  \label{eq:Gamma_def}
\Gamma_{j,j'}^{(g)} = \int\limits_\Omega d\x \, v_j^{(-g)}(\x)\, v_{j'}^{(g)}(\x) 
\end{equation}
is the overlap between two eigenfunctions $v_j^{(-g)}(\x)$ and
$v_{j'}^{(g)}(\x)$.  The integrals in Eqs. (\ref{eq:mu_def},
\ref{eq:Gamma_def}) can be computed directly by using the
representation (\ref{eq:vn_uk}):
\begin{equation}
\mu_j = \frac{1}{\sqrt{|\Omega|}} \int\limits_\Omega d\x \sum\limits_k X_{j,k} u_k(\x) = [X U]_j ,
\end{equation}
where
\begin{equation}
U_k = \frac{1}{\sqrt{|\Omega|}} \int\limits_\Omega d\x \, u_k(\x) .
\end{equation}
Since $u_k(\x)$ are orthogonal to $u_0(\x) = 1/\sqrt{|\Omega|}$ for
Neumann boundary condition, one gets
\begin{equation}
\mu_j = X_{j,0} .
\end{equation}
Similarly,
\begin{equation}   
\Gamma_{j,j'} =  \sum\limits_k X_{j,k}^* \sum\limits_{k'} X_{j',k'} \int\limits_\Omega d\x \, u_k^*(\x)\, u_{k'}(\x)
= [X^* X^\dagger]_{j,j'} .
\end{equation}

In practice, one can only construct a finite-dimensional
approximation of the infinite-dimensional matrix $\Lambda + i\bar{g}B$
by using a large but finite number $N$ of eigenmodes of the Laplace
operator (see \ref{sec:Asphere} and \ref{sec:Acapped} for details).
The numerical diagonalization of the truncated matrix of size $N\times
N$ yields $N$ eigenvalues, which are {\it expected} to converge to the
eigenvalues of the Bloch-Torrey operator as $N$ goes to infinity.
This conjecture was supported by numerical evidence: when computing a
given number of eigenvalues (and eigenfunctions) by diagonalizing
truncated matrices with larger and larger $N$, we observed that they
rapidly become almost independent of $N$.  Another indirect evidence
for this convergence comes from the fact that the eigenvalues obtained
from truncated matrices obey the large-$g$ asymptotic behavior derived
for the eigenvalues of the Bloch-Torrey operator
\cite{Grebenkov18,Moutal19}.  At the same time, a rigorous proof of
the convergence is still missing.  In fact, for many non-Hermitian
matrices, the eigenvalues are known to be very sensitive to
perturbations (such as truncation) so that the convergence may fail.
For example, the spectra of banded Toeplitz matrices of increasing
sizes do not converge to the spectrum of their limiting operator
acting on an appropriate infinite-dimensional space
\cite{Reichel92,Trefethen97,Trefethen}.  A systematic study
of the convergence presents thus an interesting perspective.

\section{Matrix elements for a sphere}
\label{sec:Asphere}

We summarize the matrix elements needed for computing the
eigenfunctions of the Bloch-Torrey operator in a sphere of radius $R$
with reflecting boundary.  For the reduced operator $\hat{\B}_g$, the
matrix representation was derived in \cite{Grebenkov07,Grebenkov08}:
\begin{equation}
\Lambda_{nk,n'k'} = \delta_{n,n'} \delta_{k,k'} \frac{\alpha_{nk}^2}{R^2} \,,
\end{equation}
where $\alpha_{nk}$ are the positive zeros of $j'_n(z)$ (with $n =
0,1,2,\ldots$), enumerated by $k = 0,1,2,\ldots$, and
\begin{equation}
\fl
B_{nk,n'k'} = \delta_{n,n'\pm 1} \frac{n+n'+1}{(2n+1)(2n'+1)} \beta_{nk} \beta_{n'k'} 
\frac{\alpha_{nk}^2 + \alpha_{n'k}^2 - n(n'+1) - n'(n+1)+1}{(\alpha_{nk}^2 - \alpha_{n'k'}^2)^2} \,,
\end{equation}
with
\begin{equation}
\beta_{nk} = \left(\frac{(2n+1)\alpha_{nk}^2}{\alpha_{nk}^2 - n(n+1)}\right)^{1/2} \,,  \qquad \beta_{00} = \sqrt{3/2}.
\end{equation}
Here we use the double index $nk$ to enumerate the elements of the
matrices $\Lambda$ and $B$.

The matrix elements of the full Bloch-Torrey operator $\B_g$ were
obtained in \cite{Ozarslan09} that we reproduce below for
completeness.  As discussed in Sec. \ref{sec:sphere}, the Laplacian
eigenfunctions $u_{nkm}$ are now enumerated by a triple index $nkm$,
with $m$ ranging from $-n$ to $n$, while the eigenvalues
$\lambda_{nkm} = \alpha_{nk}^2/R^2$ do not depend on $m$ and thus
$(2n+1)$ times degenerate.  As a consequence, the matrix $\Lambda$
takes a block-diagonal form, with the elements
\begin{equation}
\Lambda_{nkm,n'k'm'} = \delta_{m,m'} \Lambda_{nk,n'k'} = \delta_{m,m'} \delta_{n,n'} \delta_{k,k'} \frac{\alpha_{nk}^2}{R^2} \,.
\end{equation}

In turn, the matrix $B$ representing the gradient term, depends on the
direction $\e_G$ of the gradient $\G = G \e_G$.  Encoding this
direction in spherical coordinates by angles $\theta_G$ and $\phi_G$
as
\begin{equation}
\e_G = \sin\theta_G \cos\phi_G \e_x + \sin \theta_G \sin\phi_G \e_y + \cos\theta_G \e_z,
\end{equation}
one can represent this gradient by the matrix
\begin{eqnarray}  
B_{nkm,n'k'm'} & = \frac{1}{R}\int\limits_{\Omega} d\x \, [u_{nkm}(\x)]^* (\e_G \cdot \x) u_{n'k'm'}(\x) \\  \nonumber
& = \biggl[\sin\theta_G \cos\phi_G B^x + \sin \theta_G \sin\phi_G B^y + \cos\theta_G B^z\biggr]_{nkm,n'k'm'},
\end{eqnarray}
with three matrices $B^x$, $B^y$ and $B^z$, representing respectively
the operators of multiplication by $x$, $y$, and $z$ in the Laplacian
eigenbasis.

For the gradient along $z$ axis, one gets
\begin{equation}
\fl
B_{nkm,n'k'm'}^z = \delta_{m,m'} B_{nk,n'k'} \sqrt{1 - \frac{m^2}{(\max\{n,n'\})^2}}   \qquad (|m| \leq \min\{n,n'\}).
\end{equation}
Expectedly, one retrieves the matrix elements $B_{nk,n'k'}$ when $m =
m' = 0$.
For two other components, the only nonzero elements are
\begin{eqnarray}  \nonumber
B_{nkm,(n+1)k'm'}^x & = \frac{B_{nk,(n+1)k'}}{2}  \left(\delta_{m',m-1} \frac{\sqrt{(n-m+1)(n-m+2)}}{n+1} \right. \\
& \left. - \delta_{m',m+1} \frac{\sqrt{(n+m+1)(n+m+2)}}{n+1} \right)  ,\\    \nonumber
B_{nkm,(n-1)k'm'}^x & = - \frac{B_{nk,(n-1)k'}}{2}  \left(\delta_{m',m-1} \frac{\sqrt{(n+m-1)(n+m)}}{n}  \right. \\
& \left. - \delta_{m',m+1} \frac{\sqrt{(n-m-1)(n-m)}}{n} \right)  ,
\end{eqnarray}
and
\begin{eqnarray}  \nonumber
B_{nkm,(n+1)k'm'}^y & = i \frac{B_{nk,(n+1)k'}}{2}  \left(\delta_{m',m-1} \frac{\sqrt{(n-m+1)(n-m+2)}}{n+1} \right. \\
& \left. + \delta_{m',m+1} \frac{\sqrt{(n+m+1)(n+m+2)}}{n+1} \right)  ,\\  \nonumber
B_{nkm,(n-1)k'm'}^y & = - i\frac{B_{nk,(n-1)k'}}{2}  \left(\delta_{m',m-1} \frac{\sqrt{(n+m-1)(n+m)}}{n} \right. \\
& \left. + \delta_{m',m+1} \frac{\sqrt{(n-m-1)(n-m)}}{n} \right)  .
\end{eqnarray}
We also compute the elements of the matrix $W$ defined by
Eq. (\ref{eq:U_def}):
\begin{equation}  \label{eq:U_sphere}
W_{nkm,n'k'm'} = (-1)^m \delta_{n,n'} \delta_{k,k'} \delta_{m,-m'} .
\end{equation}
We stress that this matrix is not the identity.

\section{Matrix elements for a capped cylinder}
\label{sec:Acapped}

In order to study the effect of anisotropy, we consider diffusion in a
capped cylinder of radius $R$ and height $H$: $\Omega = \{ \x =
(x,y,z)\in\R^3 ~:~ x^2+y^2 < R^2, ~ -H/2 < z < H/2\}$.  Since the
lateral diffusion along the $z$ axis is independent from the
transverse diffusion in the $xy$ plane, one usually considers
separately the gradient encoding in these orthogonal directions.  For
a standard pulsed-gradient spin-echo sequence with two opposite
gradient pulses, it is therefore enough to consider two reduced
Bloch-Torrey operators: $\hat{\B}^z_g = - \partial_z^2 + igz$ on the
interval $(-H/2,H/2)$, and $\hat{\B}^{xy}_g = - (\partial_r^2 + r^{-1}
\partial_r) + ig r\cos\theta$ for a disk of radius $R$.  The matrix
elements for both operators were given explicitly in
\cite{Grebenkov08,Grebenkov07}.  The spectral properties of
$\hat{\B}^z_g$ were thoroughly investigated in
\cite{Stoller91,Grebenkov14,Grebenkov17}, while the spectrum of
$\hat{\B}^{xy}_g$ was discussed in \cite{deSwiet94,Moutal22} (see
references therein).  However, more sophisticated pulsed-gradient
sequences with several gradient directions require the knowledge of
the whole Bloch-Torrey operator $\B_g$.  We summarize the matrix
elements needed for constructing the spectrum of this operator.

The separation of variables allows one to get the eigenbasis of the
Laplacian operator explicitly in cylindrical coordinates
$(r,\theta,z)$ as
\begin{equation}  \label{eq:u_capped}
u_{nklm}(r,\theta,z) = u_{nkl}^{\rm d}(r,\theta) \, \frac{\sqrt{2-\delta_{m,0}}}{\sqrt{H}} \cos(\pi m (z+H/2)/H),
\end{equation}
where $u_{nkl}^{\rm d}(r,\theta)$ are the Laplacian eigenfunctions for
a disk of radius $R$:
\begin{equation}  \label{eq:u_disk}
u_{nkl}^{\rm d}(r,\theta) = \frac{\sqrt{2-\delta_{n,0}}}{\sqrt{\pi} R} \, \frac{\beta_{nk}}{J_n(\alpha_{nk})} \, J_n(\alpha_{nk}r/R) 
\times \left\{ \begin{array}{l l} \cos(n\theta) & (l=1), \\  \sin(n\theta) & (l=2), \\ \end{array} \right. 
\end{equation}
where $J_n(z)$ is the Bessel function of the first kind, $\alpha_{nk}$
are the positive zeros of $J'_n(z)$ enumerated by $k = 0,1,2,\ldots$,
and
\begin{equation}
\beta_{nk} = \frac{\alpha_{nk}}{\sqrt{\alpha_{nk}^2 - n^2}} \,, \qquad  \beta_{00} = 1.
\end{equation}
One sees that the Laplacian eigenfunctions $u_{nklm}$ are enumerated
by the multi-index $nklm$, with $m = 0,1,2,\ldots$.  The associated
eigenvalue is simply 
\begin{equation}
\lambda_{nklm} = \frac{\alpha_{nk}^2}{R^2} + \frac{\pi^2 m^2}{H^2} \,.  
\end{equation}
In general, the eigenvalue $\lambda_{nklm}$ is twice degenerate for $n
> 0$ and simple for $n = 0$ (in this case, $u_{0k2}^{\rm d}(r,\theta)
\equiv 0$ is not an eigenfunction and thus excluded).  However, one
can get higher-order degeneracy for specific values of the aspect
ratio $H/R$.

The structure of the Laplacian eigenfunctions allows one to construct
explicitly the matrices $\Lambda$ and $B^i$ ($i=x,y,z$) representing
the Laplace operator and the gradient along three coordinate axes.
From the practical point of view, it is convenient to construct these
matrices by reproducing their block structure:
\begin{eqnarray}
\fl
\Lambda & = \left(\begin{array}{ccccc}  
\Lambda_{\rm d} & 0 & 0 & 0 & \ldots \\ 
0 & \Lambda_{\rm d} + (\pi^2/H^2) \I & 0 & 0 & \ldots \\ 
0 & 0 & \Lambda_{\rm d} + (4\pi^2/H^2) \I & 0 & \ldots  \\ 
0& 0 & 0 & \Lambda_{\rm d} + (9\pi^2/H^2) \I & \ldots  \\
\ldots&\ldots& \ldots & \ldots & \ldots \\ \end{array} \right),  \\
\fl
B^{x,y} & = \left(\begin{array}{ccccc}  
B^{x,y}_{\rm d} & 0 & 0 & 0 & \ldots \\ 
0 & B^{x,y}_{\rm d} & 0 & 0 & \ldots \\ 
0 & 0 & B^{x,y}_{\rm d} & 0 & \ldots  \\ 
0& 0 & 0 & B^{x,y}_{\rm d} & \ldots  \\
\ldots&\ldots& \ldots & \ldots & \ldots \\ \end{array} \right),  \\  
\fl
B^{z} & =  \left(\begin{array}{ccccc}  
B^{\rm i}_{0,0} \I & B^{\rm i}_{0,1} \I & B^{\rm i}_{0,2} \I & B^{\rm i}_{0,3}\I & \ldots \\ 
B^{\rm i}_{1,0} \I & B^{\rm i}_{1,1} \I & B^{\rm i}_{1,2} \I & B^{\rm i}_{1,3}\I & \ldots \\ 
B^{\rm i}_{2,0} \I & B^{\rm i}_{2,1} \I & B^{\rm i}_{2,2} \I & B^{\rm i}_{2,3}\I & \ldots \\ 
B^{\rm i}_{3,0} \I & B^{\rm i}_{3,1} \I & B^{\rm i}_{3,2} \I & B^{\rm i}_{3,3}\I & \ldots \\ 
\ldots&\ldots& \ldots & \ldots & \ldots \\ \end{array} \right),    
\end{eqnarray}
where $\I$ is the identity matrix, $B^{\rm i}_{m,m'}$ are the matrix
elements for the interval:
\begin{equation}
\fl
B^{\rm i}_{m,m'} = ((-1)^{m+m'}-1) \sqrt{2-\delta_{m,0}} \sqrt{2-\delta_{m',0}}\, \frac{m^2 + m'^2}{\pi^2 (m^2 - m'^2)^2} \quad (m\ne m'),
\end{equation}
and $B^{\rm i}_{m,m} = 0$, with $m,m' = 0,1,2,\ldots$.  In turn,
$\Lambda_{\rm d}$ and $B_{\rm d}^{x,y}$ are the matrices representing
the Laplace operator and the gradient for the disk.  As for the case
of a sphere, there matrices were first derived explicitly in
\cite{Grebenkov07,Grebenkov08} for the reduced Bloch-Torrey operator
and then extended in \cite{Ozarslan09}.  We re-derive the extended
expressions in a slightly different form.  Skipping straightforward
computations, we get
\begin{equation}
[\Lambda_{\rm d}]_{nkl,n'k'l'} = \delta_{n,n'} \delta_{k,k'} \delta_{l,l'} \frac{\alpha_{nk}^2}{R^2} 
\end{equation}
and
\begin{equation}
[B_{\rm d}^x]_{nk1,n'k'1} = B^{\rm d}_{nk,n'k'} , \qquad [B_{\rm d}^x]_{nk2,n'k'2} = B^{\rm d}_{nk,n'k'} (1 - \delta_{n+n',1}), 
\end{equation}
\begin{equation}
[B_{\rm d}^x]_{nk1,n'k'2} = [B_{\rm d}^x]_{nk2,n'k'1} = 0,
\end{equation}
where
\begin{equation}
B^{\rm d}_{nk,n'k'} = \delta_{n,n'\pm 1} (1 + \delta_{n,0} + \delta_{n',0})^{1/2} \beta_{nk} \beta_{n'k'} 
\frac{\alpha_{nk}^2 + \alpha_{n'k'}^2 - 2nn'}{(\alpha_{nk}^2 - \alpha_{n'k'}^2)^2} 
\end{equation}
is the matrix $B^{\rm d}$ for the reduced Bloch-Torrey operator derived in
\cite{Grebenkov07,Grebenkov08}.
Similarly, one has
\begin{equation}
\fl
[B_{\rm d}^y]_{nk1,n'k'1} = [B_{\rm d}^y]_{nk2,n'k'2} = 0 , 
\end{equation}
\begin{equation}
\fl
[B_{\rm d}^y]_{nk1,(n+1)k'2} = B^{\rm d}_{nk,(n+1)k'} , \quad  
[B_{\rm d}^y]_{nk1,(n-1)k'2} = -B^{\rm d}_{nk,(n-1)k'} (1 - \delta_{n+n',1}),
\end{equation}
and
\begin{equation}
\fl
[B_{\rm d}^y]_{nk2,(n-1)k'1} = B^{\rm d}_{nk,(n-1)k'} , \quad  
[B_{\rm d}^y]_{nk2,(n+1)k'1} = -B^{\rm d}_{nk,(n+1)k'} (1 - \delta_{n+n',1}).
\end{equation}

\section{Orthogonalization of eigenfunctions with degenerate eigenvalues}
\label{sec:A_ortho}

The relation (\ref{eq:vj_orthogonality}) ensures the orthogonality of
eigenfunctions $v_j^{(g)}$ and $v_{j'}^{(g)}$ with respect to the
bilinear form $\langle \cdot,\cdot\rangle$ if two associated
eigenvalues are distinct.  In turn, if an eigenvalue is $n$ times
degenerate, i.e., there are distinct indices $j_1,\ldots,j_n$ such
that $\lambda_{j_1}^{(g)} = \lambda_{j_2}^{(g)} = \ldots =
\lambda_{j_n}^{(g)}$, the associated eigenfunctions $v_{j_i}^{(g)}$
form an eigenspace of dimension $n$, in which any $n$ linearly
independent combinations of $v_{j_i}^{(g)}$ can be chosen as
eigenfunctions.  Even though this ambiguity does not affect the
computation of the macroscopic signal via the matrix formalism, the
coefficients in the spectral expansion (\ref{eq:S_exact}) can be
sensitive to this choice.  Moreover, a proper graphical representation
of each eigenfunction and its visual interpretation require to choose
the linear combinations that respect the orthogonality.  In this
Appendix, we briefly describe a straightforward orthogonalization
procedure for the case $n = 2$.  This procedure was sufficient for the
cases of a sphere and a capped cylinder.  We also focus on the generic
setting when $g$ is not a branch point (indeed, as one eigenfunction
disappears at the branch point, the analysis of this particular
situation is more subtle, see discussion in
\cite{Moutal22}).

Let $v_j$ and $v_{j'}$ be two eigenfunctions with the same eigenvalue
(we dropped here the superscript $^{(g)}$ for brevity), and their
(non)-orthogonality is characterized by the matrix
\begin{equation}
C = \left(\begin{array}{cc} \langle v_j, v_j\rangle & \langle v_j, v_{j'}\rangle \\ 
\langle v_{j'}, v_j\rangle & \langle v_{j'}, v_{j'}\rangle \\ \end{array}\right) .
\end{equation} 
We aim at constructing two linear combinations,
\begin{equation}
\hat{v}_j = a v_j + b v_{j'}   , \qquad \hat{v}_{j'} = c v_j + d v_{j'},
\end{equation}
whose unknown coefficients $a,b,c,d$ are chosen to ensure the
orthonormality of these combinations: $\langle \hat{v}_j ,
\hat{v}_{j'}\rangle = \delta_{j,j'}$.  In a matrix form, we have
\begin{equation}
\left(\begin{array}{c} \hat{v}_j \\ \hat{v}_{j'} \\ \end{array}\right) = \left(\begin{array}{c c} a & b \\ c & d \\ \end{array}\right)
\left(\begin{array}{c} v_j \\ v_{j'} \\ \end{array}\right),
\end{equation}
so that
\begin{equation}
\left(\begin{array}{cc} \langle \hat{v}_j, \hat{v}_j\rangle & \langle \hat{v}_j, \hat{v}_{j'}\rangle \\ 
\langle \hat{v}_{j'}, \hat{v}_j\rangle & \langle \hat{v}_{j'}, \hat{v}_{j'}\rangle \\ \end{array}\right) 
= \left(\begin{array}{c c} a & b \\ c & d \\ \end{array}\right) C
\left(\begin{array}{c c} a & c \\ b & d \\ \end{array}\right).
\end{equation}
By equating the left-hand side to the identity matrix, one can
multiply this equation on the left by $\left(\begin{array}{c c} a & b
\\ c & d \\ \end{array}\right)^{-1}$ and on the right by
$\left(\begin{array}{c c} a & c \\ b & d \\
\end{array}\right)^{-1}$, to get equations on the unknown coefficients
$a,b,c,d$:
\begin{equation}  \label{eq:A_matrixeq}
\frac{1}{(ad-bc)^2} \left(\begin{array}{c c} b^2 + d^2 & -ab-cd \\ -ab-cd & a^2 + c^2 \\ \end{array}\right) = C.
\end{equation} 

To proceed, we parameterize the unknown coefficients as:
\begin{equation}
a = A^{-1} \cos\alpha, \quad b = A^{-1} \sin \alpha, \quad c = -B^{-1}\sin\alpha, \quad d = B^{-1} \cos\alpha.
\end{equation}
In the Hermitian setting, $\alpha$ could be interpreted as a rotation
angle, while $A$ and $B$ as rescaling factors.  In our case, this is a
formal representation; in particular, all three parameters $\alpha$,
$A$, $B$ can take complex values.  Substituting these expressions into
Eq. (\ref{eq:A_matrixeq}) yields three equations:
\begin{eqnarray}
A^2 \cos^2\alpha + B^2 \sin^2\alpha & = & C_{1,1}, \\
A^2 \sin^2\alpha + B^2 \cos^2\alpha & = & C_{2,2}, \\
(A^2 - B^2) \sin(2\alpha) & = & 2C_{1,2}.
\end{eqnarray}
One can solve these equations as
\begin{eqnarray}
\alpha &=& \frac12 \mathrm{atan}\left(\frac{2C_{1,2}}{C_{1,1} - C_{2,2}}\right), \\
A^2 &=& \frac{C_{1,1} \cos^2\alpha - C_{2,2} \sin^2\alpha}{\cos^2\alpha - \sin^2\alpha} , \qquad
B^2 = \frac{C_{2,2} \cos^2\alpha - C_{1,1} \sin^2\alpha}{\cos^2\alpha - \sin^2\alpha} .
\end{eqnarray}
When $C_{1,1}$ is close to $C_{2,2}$, $\alpha$ is close to $\pm
\pi/4$, so that the above expressions may be numerically unstable.  In
this case, it is more convenient to use another representation:
\begin{equation}
A^2 = \frac{C_{1,1} + C_{2,2}}{2} + \frac{C_{1,2}}{\sin(2\alpha)} , \qquad
B^2 = \frac{C_{1,1} + C_{2,2}}{2} - \frac{C_{1,2}}{\sin(2\alpha)} .
\end{equation}
In this way, we have explicit expressions for the linear
transformation from an non-orthogonal pair of eigenfunctions $v_j$ and
$v_{j'}$ to an orthonormal pair of eigenfunctions $\hat{v}_j$ and
$\hat{v}_{j'}$.

In practice, once the matrix $X$ of coefficients in
Eq. (\ref{eq:vn_uk}) is found by solving the eigenvalue problem
(\ref{eq:LambdaX}), one can evaluate the matrix $X W X^\dagger$ that
represents the orthogonality of the eigenfunctions $v_j^{(g)}$.
According to Eq. (\ref{eq:vj_vj_XUX}), this matrix should be equal to
the identity matrix.  As discussed above, this is ensured by
Eq. (\ref{eq:vj_orthogonality}) for any pair of eigenfunctions with
distinct eigenvalues.  As a consequence, nonzero off-diagonal elements
of the matrix $X W X^\dagger$ are only possible for pairs of
eigenfunctions with the same eigenvalue.  One can therefore search for
such nonzero off-diagonal elements and apply the above
orthonormalization procedure for each such pair.  This procedure was
applied for most spectral computations in this work.

\end{document}